\documentclass[]{interact}

\usepackage{epstopdf}
\usepackage{subfigure}

\usepackage{natbib}
\bibpunct[, ]{(}{)}{;}{a}{}{,}
\renewcommand\bibfont{\fontsize{10}{12}\selectfont}

\usepackage{hyperref}
\usepackage[T1]{fontenc}
\usepackage{float}
\usepackage{anyfontsize}

\newcommand{\pr}{\mathbb{P}}

\newcommand{\aipw}{\mathrm{AIPW}}

\newcommand{\amw}{\mathrm{AMW}}

\newcommand{\reg}{\mathrm{reg}}
\newcommand{\mat}{\mathrm{mat}}

\newcommand{\J}{\mathcal{J}}

\newtheorem{lem}{Lemma}
\newtheorem{thm}{Theorem}

\newtheorem{assumption}{Assumption}

\begin{document}

\articletype{ORIGINAL ARTICLE}

\title{[DISS]Augmented Match Weighted Estimators: New Methods for Estimating Average Treatment Effects Under Extreme Propensity Scores}

\author{
\name{Tanchumin Xu\textsuperscript{a,b}, Yunshu Zhang\textsuperscript{a,c*}\thanks{*Corresponding author: Yunshu Zhang. Email: yunshuzhang18@gmail.com}, Shu Yang\textsuperscript{a}}
\affil{\textsuperscript{a}Department of Statistics, North Carolina State University, North Carolina, U.S.A.; \textsuperscript{b}Bioinformatics Research Center, North Carolina State University, North Carolina, U.S.A.; \textsuperscript{c}Department of Biostatistics, University of Pennsylvania, Pennsylvania, U.S.A.}
}

\maketitle

\begin{abstract}
Propensity score matching (PSM) and augmented inverse propensity weighting (AIPW) are used in observational studies to estimate causal effects. The AIPW estimator is doubly robust and locally efficient but can be unstable when the propensity scores are close to zero or one. PSM circumvents the instability of propensity score weighting but it hinges on the correctness of the propensity score model and cannot attain the semiparametric efficiency bound. The fixed number of matches, $K$, renders PSM nonsmooth and thus invalidates standard bootstrap inference. 

This article presents novel augmented match weighted (AMW) estimators that combine the advantages of matching and weighting estimators. AMW adheres to the form of AIPW for its double robustness and local efficiency but it mitigates the instability. We replace inverse propensity weights with matching weights resulting from PSM with unfixed $K$. Meanwhile, we propose a new cross-validation procedure to select $K$ that minimizes the mean squared error anchored around an unbiased estimator of the causal estimand. We derive the limiting distribution showing AMW estimators enjoy the double robustness property and can achieve the semiparametric efficiency bound if both nuisance models are correct. As a byproduct of unfixed $K$ which smooths AMW estimators, nonparametric bootstrap can be adopted for variance estimation. Furthermore, simulation and real-data applications support that the AMW estimators are stable and their variances can be obtained by nonparametric bootstrap.
\end{abstract}


\section{Introduction}\label{sec:intro}

Causal inference has garnered significant attention in recent years
due to its potential applications in various fields, including sociology,
economics, and medicine. Causal inference aims to identify the cause-and-effect
relationships between variables of interest. In observational studies,
it is often crucial to estimate the average treatment effect (ATE)
and the average treatment effect on the treated (ATT). However, this
can be challenging since treatment assignment may be correlated with
various covariates associated with potential outcomes. Moreover, due
to the non-random assignment of treatments, there may be an imbalance in both treatment and control groups \citep{imbens2015causal}. These
challenges arise because researchers in observational studies have
no control over the treatment assignment to units. As a result, causal
inference methods are employed to adjust for confounding variables
and estimate the causal effects of treatments.

Matching \citep{cochran1973controlling,rubin1973matching} has a rich
history for conducting causal inference in observational studies.
These methods involve pairing each treated unit to untreated units
that have similar covariates, aiming to remove the confounding bias
and balance the two groups \citep{stuart2010matching}. Intuitively,
the nearest neighbor (NN) matching approach matches based on covariates
$X$, with a fixed number of matches, $K=1$. Some approaches also
allow $K$ to change with varying sample sizes.  \citet{heckman1997matching}
applied kernel matching and local linear kernel matching estimators
conditioned on covariates $X$, which reduced the confounding bias
substantially.  Note that by allowing $K$
to diverge with different samples, the K-nearest neighbor (KNN) matching
estimator with covariates $X$ can be doubly robust and semiparametrically
efficient. However, due to the curse of dimensionality, matching on
the full set of covariates may not entirely remove the confounding
bias in high-dimensional settings. To overcome this challenge, several dimension reduction techniques
have been developed. \citet{rosenbaum1983central} illustrated the central
role of the propensity score to balance the two groups and hence proposed
propensity score matching (PSM) to estimate causal effects. PSM with
a fixed number of matches $K$ has been extensively studied \citep{abadie2006large,abadie2012martingale,abadie2016matching,stuart2010matching}.
However, PSM is only singly robust, as it yields consistent estimators
only with the correct propensity score models. \citet{leacy2014joint}
and \citet{yang2020multiply} proposed doubly robust matching estimators
and explored their asymptotic properties. These estimators are consistent
for causal effects if either a model for the propensity score or a
model for the outcomes is correctly specified. However, the matching
estimators with a fixed $K$ are inefficient, and the nonparametric
bootstrap is not amenable to estimating their corresponding variances.
The failure of bootstrap is because the matching estimators
with a fixed $K$ lack smoothness, and the distribution of the number
of times when each unit is utilized as a match cannot be replicated
by the bootstrap procedure\citep{abadie2008failure}. To address this
issue, \citet{otsu2016bootstrap} developed a weighted bootstrap approach
by resampling based on certain linear forms of the estimators when
matching on $X$ with a fixed number of matches. They demonstrated
the validity of their approach, but it did not extend to the PSM estimator.

In addition to matching, several other ways exist to obtain valid
estimations for treatment effects. The inverse probability of weighting
(IPW) \citep{imbens2015causal} estimator addresses the confounding
issue by assigning the inverse of the propensity scores as weights
to all units. But it is only singly robust and may be unstable even
when the propensity score model is correctly specified. Besides, \citet{robins1994estimation,bang2005doubly,cao2009improving} provided doubly robust weighting estimators,
i.e., the augmented IPW (AIPW) estimators, which achieve the semiparametric
efficiency bound \citep{hahn1998role,tsiatis2007semiparametric}.
However, such weighting estimators can still be unstable when the
propensity scores are close to one or zero.

In this paper, we provide a comprehensive review of the PSM and AIPW
estimators and revisit their strengths and weaknesses. Matching has
advantages over weighting: first, and most significantly, matching
does not involve the potentially unstable inverse of propensity scores
\citep{frolich2004finite} and second, matching is intuitively appealing
to replicate a randomized experiment \citep{heckman1997matching,dehejia&wahba98,rubin2006matched,stuart2010matching}.
However, AIPW is easier to implement and can achieve the semiparametric
efficiency bound.

Motivated by the distinguishing features of PSM and AIPW, we propose
a new type of estimator called the augmented match weighted (AMW)
estimator, which is stable, efficient, and doubly robust for estimating
the ATE and ATT. The AMW estimator introduces a fresh perspective
to reveal the connections between matching and weighting. On the one
hand, they can be viewed as an augmentation of PSM estimators with
outcome models. On the other hand, it is connected to the AIPW estimators
by replacing inverse propensity score weights with matching weights
from PSM. Hence, the AMW estimators can be seen as a combination of
the PSM and AIPW estimators. In addition, the AMW estimator has a
smooth property, as the number of matches is unfixed, and the standard
bootstrap is valid for estimating the variance. The AMW estimator
has three key advantages. Firstly, it is stable, unlike the weighting
type estimators. The kernel estimations for weights are less extreme
than the propensity scores from parametric models, making the AMW
estimator more stable when propensity scores are extreme. Secondly,
if both nuisance models are correct, the AMW estimator achieves the
semiparametric efficiency bound. Thirdly, the AMW estimator enjoys
the double robustness property, which means it is still consistent
when either the propensity score model or outcome model is misspecified.

Besides, selecting the tuning parameter $K$ is an important aspect
of the AMW estimator. \citet{lin2023estimation} suggested allowing $K$ to increase with the sample size and showed that doing so can improve the asymptotic properties of matching estimators from a theoretical perspective. In practice, however, data-driven procedures are required to determine an appropriate $K$ for a given sample size. Cross-validation (CV) is commonly used
for model selection in causal inference. \citet{cui2019selective}
minimized the CV-based pseudo-risk for doubly robust, semiparametric
estimating functions from the nuisance models. CV can also be used
to choose different estimators for conditional treatment effects \citep{rolling2014model}.  \citet{brookhart2006semiparametric} and  \citet{rothenhausler2020model}
developed CV approaches to select tuning parameters based on benchmark
estimators using baseline parameters. For example, \citet{ju2019adaptive}
optimized a closed-form mean squared error (MSE) based on the sum of
the bias and variance terms to select cutoff parameters for the propensity
score truncation problem where the IPW estimator truncated at $0th$
percentile is treated as the reference estimator. To select the unfixed
$K$ for the AMW estimator, we propose a new CV approach in practice.
The idea is to treat the AMW estimator with $K=1$ as an unbiased
benchmark estimator and study the bias of the AMW estimator with candidate
parameter $K$. The variance can be estimated by naive bootstrap,
and we can minimize the MSE for an appropriate parameter $K$.

We present the theoretical properties in two steps. In the first step,
we study their large sample distributions with known parameters under
the KNN framework by \citet{mack1979multivariate} and \citet{yangkim2016}.
We derive their asymptotic linear expansions for AMW and show that
they have approximately normal distributions. The asymptotic variances
indicate that the AMW estimators attain the semiparametric efficiency
bounds when the propensity score and outcome models are correct. In
the second step, we extend the results of  \citet{andreou2012alternative}
and \citet{abadie2016matching} to derive the large sample properties
of the AMW estimators when nuisance parameters are estimated. Then
we approximate their limiting normal distributions based on the estimated
scores, building on the works of \citet{abadie2016matching} and  \citet{yang2020multiply}.

The article is organized as follows. Section \ref{sec:Model-procedure}
provides a basic setup and briefly reviews the matching and AIPW estimators
to motivate the proposed estimators. Section \ref{sec:Proposed-methodology}
gives a basic framework and algorithms to construct the AMW estimators
and corresponding variance estimators. In Section \ref{sec:Main-theory},
main theories are established for the AMW estimators with both known
and estimated scores. Section \ref{sec:Simulation-study} reports
simulation studies to investigate the performances of different estimators. Section \ref{sec:Real-data-analysis}
showcases the proposed approach through three real data applications.
Section \ref{sec:Discussion} concludes with a discussion.

\section{A review of methodologies} \label{sec:Model-procedure}

\subsection{Basic setup}

We follow the potential outcomes framework throughout this paper.
Assume $\{X_{i},A_{i},Y_{i}(0),\allowbreak Y_{i}(1)\}_{i=1}^{n}\stackrel{{\rm i.i.d.}}{\sim}\{X,A,Y(0),Y(1)\}$
in the observational study, and let $A_{i}$ be the binary treatment,
$X_{i}$ be the pre-treatment covariates and $Y_{i}(a)$ be the potential
outcome conditioned on the treatment $a$ ($a=0,1$). Given the standard
Stable Unit Treatment Value Assumption \citep{rosenbaum1983central},
the observed outcome is $Y_{i}=A_{i}Y_{i}(1)+(1-A_{i})Y_{i}(0)$.
Since $\{X_{i},A_{i},Y_{i}(0),Y_{i}(1)\}_{i=1}^{n}$ are i.i.d, the
observations $\{X_{i},A_{i},Y_{i}\}_{i=1}^{n}$ are also i.i.d. The
sample size is denoted as $n$, while $n_{1}$ and $n_{0}$ represent
the sizes of the treated and control subpopulation, respectively.
This paper focus on two commonly-used estimands: the ATE $\tau=\mathbb{E}\{Y(1)-Y(0)\}$
and the ATT $\tau^{t}=\mathbb{E}\{Y(1)-Y(0)\mid A=1\}$. To simplify
the presentation, for a generic variable $V$, denote $u_{a}(V)=\mathbb{E}\{Y(a)\mid V\},\,\sigma_{a}^{2}(V)=\mathbb{V}\{Y(a)|V\},\,e(V)=\mathbb{P}(A=1\mid V),$ where $u_{a}(V)$ represents the mean function for the outcome, $\sigma_{a}^{2}(V)$
represents the variance function, and $e(V)$ denotes the mean function
for the treatment. When $V$ corresponds to the pre-treatment covariates
$X$, $e(X)$ is commonly referred to as the propensity score.

The basic framework relies on several fundamental assumptions:

\begin{assumption}\label{assumption1}

There exist two constants $c_{0}$ and $c_{1}$ such that $0<c_{0}\leq e(X)\leq c_{1}<1$ almost surely. $\{Y(0),Y(1)\}\perp A\mid X$.

\end{assumption}

The first part for Assumption \ref{assumption1} implies that all units can be allocated
to either the treatment or control group, indicating a considerable
overlap in the distributions of pretreatment variables. The second part is also referred to as strong ignorability of treatment
assignment, suggesting that treatments and potential outcomes are
independent given adequate covariates. While researchers cannot test
this assumption, they can enhance the reliability of their inferences
by collecting more covariates.

\subsection{K-nearest neighbor matching}

In order to explain the concept of AMW estimators, we will review
the class of matching estimators. To fix ideas, we first consider
the case of matching with replacement, where the number of matches
to be fixed as $K$ ($K\geq1$). This approach is equivalent to the
$K$-nearest neighbor (KNN) approach (Here, KNN is the conventional
name, so we will use $K$ in our context). Generally, it is recommended
to use $K=1$, namely the nearest neighbor (NN) approach \citep{rubin1973matching,mack1979multivariate}
to gain fewer biases. The matching estimators with fixed $K$ are one
of the most commonly used estimators in observational studies, and
their statistical properties have been extensively studied in a series
of works \citep{abadie2006large,abadie2011bias,abadie2016matching}.
Still, we will later propose an algorithm to specify the matching
estimators with unfixed $K$ in Section \ref{subsec:unfixedK}. Based
on the matching variable $V$, let $\J_{V}(i)$ denote the set of
indices for the nearest $K$ neighbors of unit $i$ in the opposite
group. $Y_{i}(A_{i})$ is observed, the counterfactual outcome $Y_{i}(1-A_{i})$
is missing but can be estimated by averaging the observed outcomes
of the $K$ matched units in the corresponding group. Specifically, the potential
outcomes for the unit $i$ can be imputed as 
\[
\hat{Y}_{i}(1)=\begin{cases}
K^{-1}\sum_{j\in\J_{V}(i)}Y_{j} & \text{if }A_{i}=0,\\
Y_{i} & \text{if }A_{i}=1,
\end{cases};\:\hat{Y}_{i}(0)=\begin{cases}
Y_{i} & \text{if }A_{i}=0,\\
K^{-1}\sum_{j\in\J_{V}(i)}Y_{j} & \text{if }A_{i}=1.
\end{cases}
\]
Let $M_{V,i}=\sum_{l}I_{(i\in\mathcal{J}_{V}(l))}$ be the number of times that unit $i$ is being matched to other units. 
We use the Euclidean distance for matching, $||\cdot||$, although
our discussion can be applied to other distance measures. A simple matching
estimator $\hat{\tau}_{\mat}^{(0)}$ for ATE can be given:
\begin{equation}
\hat{\tau}_{\mat}^{(0)}=\frac{1}{n}\sum_{i=1}^{n}\hat{Y}_{i}(1)-\frac{1}{n}\sum_{i=1}^{n}\hat{Y}_{i}(0)=\frac{1}{n}\sum_{i=1}^{n}(2A_{i}-1)(1+K^{-1}M_{V,i})\hat{Y}_{i}.\label{eq:atematch}
\end{equation}
Then $\hat{\tau}_{\mat}^{t}$ for ATT is
\begin{equation}
\hat{\tau}_{\mat}^{t,(0)}=\frac{1}{n_{1}}\sum_{i=1}^{n}Y_{i}A_{i}-\frac{1}{n_{1}}\sum_{i=1}^{n}\hat{Y}_{i}(0)A_{i}=\frac{1}{n_{1}}\sum_{i=1}^{n}\{A_{i}-(1-A_{i})K^{-1}M_{V,i}\}\hat{Y}_{i}.\label{eq:attmatch}
\end{equation}

Matching has the advantage of producing intuitive estimators. It is
also straightforward to evalute the matching performance by examining
the balance of the covariates. A small difference between the treatment
and control groups may indicate a successful matching process. Nevertheless,
one of the drawbacks of matching is that the estimators can be biased
due to matching discrepancies. \citet{abadie2006large} showed that the
asymptotic bias has an order of $O_{\pr}(n^{1/2-1/d}),$
where $d$ represents the dimension of the matching variable $V$.
For example, when matching is directly based on the pre-treatment
covariates (i.e., $V=X$), the bias $\hat{B}_{n}$ is considerable unless the dimension
of $X$ is one. A bias-corrected matching estimator is given as $\hat{\tau}_{\mat}=\hat{\tau}_{\mat}^{(0)}-\hat{B}_{n}$.
The term $\hat{u}_{a}(V_{i})$ can be obtained either parametrically
(e.g., by a linear regression estimator), or nonparametrically. Equivalently, we update the imputed potential outcomes for unit $i$ as 
\[
\widetilde{Y}_{i}(a)=\begin{cases}
K^{-1}\sum_{j\in\J_{V}(i)}\left\{ Y_{j}+\hat{u}_{a}\left(V_{i}\right)-\hat{u}_{a}\left(V_{j}\right)\right\} & \text{if }A_{i}\neq a,\\
Y_{i} & \text{if }A_{i}=a.
\end{cases};
\]
Then the bias-corrected matching estimator is
\begin{equation}
\hat{\tau}_{\mat}=\frac{1}{n}\sum_{i=1}^{n}\{\widetilde{Y}_{i}(1)-\widetilde{Y}_{i}(0)\}.\label{eq:mat-biasc}
\end{equation}


However, correcting the matching bias by relying solely on the pre-treatment
covariates $X$ may not always remove all sources of confounding bias
due to the curse of dimensionality. As a result, it is necessary to
use sufficient statistics as the matching elements. \citet{rosenbaum1983central}
demonstrated the central role of the score $e(X)$ as a balancing
score:
\begin{lem}
Under Assumptions \ref{assumption1}, $\{Y(1),Y(0)\}\perp A\mid e(X)$,
which implies that $\tau=\mathbb{E}[\mathbb{E}\{Y\mid A=1,e(X)\}-\mathbb{E}\{Y\mid A=0,e(X)\}]$
and $\tau^{t}=\mathbb{E}[\mathbb{E}\{Y\mid A=1,e(X)\}-\mathbb{E}\{Y\mid A=0,e(X)\}|A=1]$. 
\label{lem:1}
\end{lem}
Lemma \ref{lem:1} suggests that the scores can reduce the dimension of the matching variables. When the lemma holds true, PSM controls for $e(X)$ instead of $X$ to estimate causal effects $\tau$ and $\tau^{t}$ via matching
the treatment and control subjects with similar scores. Under a correct
model, the PSM estimator is consistent and can sufficiently
eliminate biases associated with a non-random treatment assignment.
However, PSM suffers from bias due to possible model misspecification,
and it may be biased if the model is incorrect. This limitation
motivates us to improve PSM's robustness by constructing our AMW estimator
in Section \ref{sec:Proposed-methodology}. 

Before proceeding, we would like to point out that the bias-corrected
matching estimator, which is derived from the pre-treatment covariates
$X$, can be rewritten in the following form, as shown in Lemma 3.1 from \cite{lin2023estimation}:
\begin{equation}
\begin{aligned}\hat{\tau}_{\mat,X} & =\hat{\tau}_{\reg}+\frac{1}{n}\sum_{i=1}^{n}\left\{ A_{i}\left(1+\frac{M_{X,i}}{K}\right)\hat{R}_{i}-(1-A_{i})\left(1+\frac{M_{X,i}}{K}\right)\hat{R}_{i}\right\},
\end{aligned}
\label{eq:matX}
\end{equation}
where $\hat{\tau}_{\reg}=n^{-1}\text{\ensuremath{\sum_{i=1}^{n}}}\left\{ \hat{u}_{1}(X_{i})-\hat{u}_{0}(X_{i})\right\} $
is the outcome regression estimator of $\tau$, and $\hat{R}_{i}=Y_{i}-\hat{u}_{A_{i}}(X_{i})$
is the residual of unit $i$. Specifically, the bias-corrected matching
estimator based on the score $e(X)$ can also be expressed similarly 
\begin{equation}
\begin{aligned}\hat{\tau}_{\mat,ps} & =\text{\ensuremath{\frac{1}{n}\sum_{i=1}^{n}}}\left[\hat{u}_{1}\left\{ e(X_{i})\right\} -\hat{u}_{0}\left\{ e(X_{i})\right\} \right]\\
 & \,+\frac{1}{n}\sum_{i=1}^{n}\left\{ A_{i}\left(1+\frac{M_{e(X),i}}{K}\right)\hat{R}_{i}-(1-A_{i})\left(1+\frac{M_{e(X),i}}{K}\right)\hat{R}_{i}\right\} .
\end{aligned}
\label{eq:matps}
\end{equation}

Note that the first term on the right-hand side is not equivalent
to $\hat{\tau}_{\reg}$. This is why PSM cannot benefit from the double
robustness property. In the section \ref{subsec:aipw}, we
will contrast this form with the AIPW estimator and utilize these
concepts to build our AMW estimator.

\subsection{Augmented inverse probability weighting (AIPW) estimator\label{subsec:aipw}}
To enhance the effectiveness and robustness of prior matching estimators, we aim
to review some double robustness efficient estimators to gain insights. In practice,
to address the weakness of unknown score and outcome models, researchers
have been developing more robust estimators \citep{yang2020multiply,han2013estimation}.
Among them, the augmented inverse probability weighting (AIPW) estimator
\citep{robins1994estimation} is most commonly used because of its double
robustness and semiparametric efficiency. Let $\hat{u}_{a}(X)$ and
$\hat{e}(X)$ be some parametric estimators for the outcome $u_{a}(X)$
and propensity score $e(X)$. The AIPW estimator for the ATE is considered
as follows and can be rewritten as:
\begin{equation}
\begin{aligned}\hat{\tau}_{\aipw} & =\text{\ensuremath{\frac{1}{n}\sum_{i=1}^{n}}}\left[\frac{A_{i}Y_{i}}{\hat{e}(X_{i})}-\frac{(1-A_{i})Y_{i}}{1-\hat{e}(X_{i})}-\frac{\left\{ A_{i}-\hat{e}(X_{i})\right\} }{\hat{e}(X_{i})}\hat{u}_{1}(X_{i})-\frac{\left\{ A_{i}-\hat{e}(X_{i})\right\} }{1-\hat{e}(X_{i})}\hat{u}_{0}(X_{i})\right]\\
 & =\hat{\tau}_{\reg}+\frac{1}{n}\sum_{i=1}^{n}\left\{ \frac{A_{i}\hat{R}_{i}}{\hat{e}(X_{i})}-\frac{(1-A_{i})\hat{R}_{i}}{1-\hat{e}(X_{i})}\right\} .
\end{aligned}
\label{eq:AIPW}
\end{equation}
Note that (\ref{eq:AIPW}) bears resemblance to (\ref{eq:matX}),
as they both utilize the outcome regression term with some adjustments
based on the residuals $\hat{R}_{i}$. This results in the doubly robust
property. The AIPW method converges satisfactorily when both models are sup-norm consistent, or when one model is accurate enough to compensate for any inaccuracies in the other model \citep{chernozhukov2018double}. Cross-fitting with minor modification of AIPW form can be employed to enable the AIPW to achieve a first-order equivalence with the oracle AIPW \citep{newey2018cross}. However, the AIPW estimator may be unstable when the inverse
propensity scores take values near zero or one \citep{kang2007demystifying,guo2014propensity}.
Conversely, matching estimators avoid the limitation of extreme propensity
scores by employing $1+M_{X}/K$ in lieu of the inverse probabilities \citep{ding2024first}.
To conclude, we summarize:
\begin{itemize}
\item The matching estimator based on the pre-treatment covariates in (\ref{eq:matX})
is robust to extreme propensity scores, but matching on $X$ suffers
from the curse of dimensionality;
\item The PSM estimator with fixed $K=1$ in (\ref{eq:matps}) is asymptotically
unbiased in high dimensional settings and robust to extreme propensity
scores, but it suffers from model misspecifications and is inefficient;
\item The AIPW estimator in (\ref{eq:AIPW}) is doubly robust and efficient,
but the inverse probabilities suffer from extreme propensity scores.
\end{itemize}
Given these considerations, we are ready to leverage the respective
advantages of the matching estimator and the AIPW estimator to construct
the AMW estimator in the following section. Before proceeding, we
also intend to rephrase the AIPW estimator for the ATT:

\begin{equation}
\hat{\tau}_{\aipw}^{t}=
\hat{\tau}_{\reg}^{t}+\frac{1}{n_{1}}\sum_{i=1}^{n}\left\{ A_{i}\hat{R}_{i}-\frac{(1-A_{i})\hat{e}(X_{i})\hat{R}_{i}}{1-\hat{e}(X_{i})}\right\} .
\label{eq:AIPWatt}
\end{equation}

\section{Proposed methodology }\label{sec:Proposed-methodology}
\subsection{AMW estimators}
Motivated by (\ref{eq:matX})-(\ref{eq:AIPW}), we propose the AMW estimators based on the advantages of matching and weighting:
\begin{equation}
\hat{\tau}_{\amw}=\hat{\tau}_{\reg}+\frac{1}{n}\sum_{i=1}^{n}\left[A_{i}\left\{ 1+\frac{M_{e(X),i}}{K}\right\} \hat{R}_{i}-(1-A_{i})\left\{ 1+\frac{M_{e(X),i}}{K}\right\} \hat{R}{}_{i}\right],\label{eq:amw}
\end{equation}
\begin{equation}
\hat{\tau}_{\amw}^{t}=\hat{\tau}_{\reg}^{t}+\frac{1}{n_{1}}\sum_{i=1}^{n}\left[A_{i}\hat{R}_{i}-(1-A_{i})\left\{ 1+\frac{M_{e(X),i}}{K}\right\} \hat{R}{}_{i}\right].\label{eq:amwt}
\end{equation}

\citet{kang2007demystifying} found that weighting type estimators
tend to have high variability due to the denominator of the weights
being close to zero. To address this issue, AMW uses the K-nearest
neighbor method to estimate the propensity score, treating $\{1+K^{-1}M_{e(X),i}\}^{-1}$
as an estimator for the propensity score. Matching, which avoids inverting
the estimated probability of treatment, makes the resulting AMW estimator
more stable than the AIPW estimator. In this way, $1+K^{-1}M_{e(X),i}$
in (\ref{eq:amw}) serves the same purpose as\textbf{ $\{P(A=A_{i}\mid X_{i})\}^{-1}$
}in (\ref{eq:AIPW}). Thus, AMW can be viewed as an alternative version
of AIPW. On the other hand, the regression estimator $\hat{\tau}_{\reg}$
in the AMW estimator acts as a bias correction term for the PSM estimator
in (\ref{eq:matps}), bringing protection against incorrect modeling
of the propensity score model. As a result, the AMW estimator is doubly
robust in the large sample setting, making it more reliable than the
PSM estimator when the propensity score model is misspecified and
more robust than $\hat{\tau}_{\reg}$ when the outcome model is misspecified.
Similarly, $K^{-1}M_{e(X),i}$ in (\ref{eq:amwt}) serves the same
role as $\{P(A=1\mid X_{i})\}\times\{P(A=0\mid X_{i})\}^{-1}$. The ATT estimator $\hat{\tau}_{\amw}^{t}$ in
(\ref{eq:amwt}) shares the same structures and properties as $\hat{\tau}_{\amw}$.

The matching type estimator with a fixed number of matches is not
smooth, and consequently, the nonparametric bootstrap \citep{efron1979bootstrap}
is unsuitable for estimating variances. To solve this problem, we
unfix the number of matches $K$ by letting $K$
increase as the sample size grows. We provide the main algorithm to obtain $\hat{\tau}_{AMW}$ and its
corresponding variance estimator. 
\begin{description}
\item [{AMW-Step1}] Posit models to obtain the propensity score $\hat{e}(X)$
and the regression estimator $\hat{\tau}_{reg}$.
\item [{AMW-Step2}] Implement Cross Validation (Section \ref{subsec:unfixedK})
to determine the optimal value of $K$ for the AMW estimators. 
\item [{AMW-Step3}] For each unit $i$ with treatment $A_{i}$, compute
$\mid\hat{e}(X_{i'})-\hat{e}(X_{i})\mid_{2}^{2}$ for $\forall i'$,
choose first $K$ smallest distances to get indexes, and calculate
$M_{\hat{e}(X),i}$ as the number of times that unit $i$ is involved
in one-time matching.
\item [{AMW-Step4}] Construct $\hat{\tau}_{AMW}$ based on $K$, $M_{\hat{e}(X),i}$
and $\hat{\tau}_{reg}$. Estimate variance for $\hat{\tau}_{AMW}$
by naive bootstrap.
\end{description}

\subsection{Unfix the number of matches in AMW estimators\label{subsec:unfixedK}}

Cross Validation (CV) is a resampling technique to select a model
for a given predictive modeling problem. Initially, the data is randomly
partitioned into training and testing subsets, and candidate models
are fitted to the training data. Then, evaluate the prediction performance
of the model via mean squared error (MSE), which compares prediction
results with test data. Lastly, choose a candidate model with minimal
MSE to strike a balance between significant variance from overfitting
and considerable bias from underfitting. Although CV is highly accurate
for prediction tasks, the selection of tuning parameters may not require
predictive ability in causal inference \citep{rothenhausler2020model}.
In this section, we propose a new form for the bias in the MSE for matching and discuss
the consistency property of the estimated $K$:
\begin{equation}
MSE\left\{ \hat{\tau}_{\amw}(K)\right\} =Var\left\{ \hat{\tau}_{\amw}(K)\right\} +Bias^{2}\left\{ \hat{\tau}_{\amw}(K)\right\} ,\label{eq:amw_mse}
\end{equation}
The AMW estimator comprises two parts: a function $B(K)$ that depends
on the number of matches $K$ and a constant term $C$, where 
\[
B(K)=\frac{\sum_{i=1}^{n}\left[A_{i}\left\{ M_{e(X),i}/K\right\} \hat{R}_{i}-(1-A_{i})\left\{ M_{e(X),i}/K\right\} \hat{R}{}_{i}\right]}{n}.
\]
The bias of AMW estimators can be represented by $B(K)$, given that $\mathbb{E}(C)=\tau$. To illustrate the relationship between $K$ and $B(K),$ we assume
the propensity scores for unit $i$ is fixed as $e_{i}^{*}$, and
the residual term is fixed as $R_{i}^{*}(a_{i})$. 
For simplicity, $\hat{m}_{a_{i}}(e_{i}^{*})$ is denoted as $
K^{-1}\sum_{j\in\mathcal{J}_{K}(i)}D_{R_{e_{i}}}\left({e}_{i}^{*}-{e}_{j}^{*}\right){R}_{j}^{*}(a_{i})I_{(A_{j}=1-a_{i})}$, where $D_{d}(x)=\frac{1}{d}D\left(\frac{x}{d}\right)$ , and $D(x)=I_{(\|x\|\leq1)}.$
Take bandwidth $d=R_{e_{i}}$, which is the random distance between
${e}_{i}^{*}$ and its furthest element among the $K$ nearest neighbors.

Let $\ensuremath{f(e_{i}^{*})}$ be density function of $e_{i}^{*}$.
We apply theorems from \citet{mack1979multivariate} to study the
bias and variance of $B(K)$ by Lemma 2.
\begin{lem}
Under regularity conditions, as $n\rightarrow\infty$, $K\rightarrow\infty$,
$K/n\rightarrow0$, we have
\begin{equation}
n^{2}\times Var\left\{ B(K)\right\} =\frac{n_0\mathbb{{V}}\left\{ AR^{*}(1)|e^{*}\right\} +n_1\mathbb{{V}}\{(1-A)R^{*}(0)|e^{*}\}}{K},\label{eq:var_amw}
\end{equation}
\begin{equation}
\resizebox{0.91\hsize}{!}{%
$Bias\{B(K)\}=\frac{K^{2}}{24n\ensuremath{f^{3}(e^{*})}}\left[\left\{ (m_{1}f)''(e^{*})-m_{1}(e^{*})f''(e^{*})\right\}\frac{n_0}{n^2_1} -\left\{ (m_{0}f)''(e^{*})-m_{0}(e^{*})f''(e^{*})\right\}\frac{n_1}{n^2_0}\right],$%
}\label{eq:bias_amw}
\end{equation}
where $(m_{a}f)(e^{*})=m_{a}\left\{f(e^{*})\right\}$.
\end{lem}
 Choosing multiple matches from the opposite groups can introduce bias
for the unit, as the second, third, and fourth closest matches are
further away from the nearest neighbors \citep{stuart2010matching}.
However, obtaining the theoretical value of the optimal $K$ is difficult in practice because Equation (\ref{eq:var_amw}) and (\ref{eq:bias_amw}) have complex forms. Hence, we propose a new data-driven CV algorithm to choose $K$
to minimize the MSE when constructing the AMW estimator. The idea is motivated from the relationship between $Bias\left\{ B(K)\right\} $
and $K$, where the bias will increase at a rate of $K^{2}$, as shown from Equation (\ref{eq:bias_amw}). If $K>\sqrt{n}$, the bias cannot be ignored even with large $n$. Intuitively,
the AMW estimator has the smallest bias when $K=1$. For a set of
candidate parameters $K^{(1)},...,K^{(p)}$, we set $K^{(1)}=1$ and
treat $\hat{\tau}_{\amw}(K^{(1)})$ as unbiased. The $Bias\{\hat{\tau}_{\amw}(K^{(j)})\}$
is calculated as the difference between $\hat{\tau}_{\amw}(K^{(j)})$
and $\hat{\tau}_{\amw}(K^{(1)})$. Specially, $Bias\{\hat{\tau}_{\amw}(K^{(1)})\}=0$
and reducing $K$ can help to decrease the bias of the AMW estimator. 

Conversely, the variance of $B(K)$ in (\ref{eq:var_amw}) decreases
as $K$ increases, indicating that a larger $K$ can reduce variance.
Thus, a trade-off problem exists between the variance and bias when
constructing the AMW estimator and estimating appreciation $K$ to
gain a balance between these factors. Besides, we propose a new data-driven CV algorithm to choose $K$
to minimize the MSE when constructing the AMW estimator.
\begin{description}
\item [{CV-Step1}] For the $j^{th}$ candidate parameter $K^{(j)}$, compute
$Var\{\hat{\tau}_{AMW}(K^{(j)})\}$ by naive bootstrap. 
\item [{CV-Step2}] Split the dataset randomly into two equal halves, compute
$\hat{\tau}_{AMW}(K^{(j)})_{1}$ for one half, and $\hat{\tau}_{AMW}(K^{(1)})_{2}$
for the rest. Obtain bias as $\hat{\tau}_{AMW}(K^{(j)})_{1}-\hat{\tau}_{AMW}(K^{(1)})_{2}$.
\item [{CV-Step3}] Repeat CV-Step2 multiple times to obtain several biases
and then take the average of these estimates to obtain a robust estimate
of the bias $Bias\{\hat{\tau}_{\amw}(K^{(j)})\}$. Compute the MSE
by adding the variance of the estimate to the square of the bias.
\item [{CV-Step4}] Select the value of $K$ that has the smallest MSE among
all options.
\end{description}
 Similarly, the algorithms to get $\hat{\tau}_{AMW}^{t}$ and its variance estimator can mimic the above strategies.

 \section{Main theory }\label{sec:Main-theory}

This section focuses on investigating the asymptotic properties of
$\hat{\tau}_{\amw}$ and $\hat{\tau}_{\amw}^{t}$. Suppose that $(X_{i},A_{i},Y_{i})$
is independently observed from $\mathbb{P}_{\theta}$, indexed by
$\theta^{\top}=(\alpha^{\top},\beta_{0}^{\top},\beta_{1}^{\top})$,
where $\alpha$ refers to the propensity score model parameter, and
$\beta_{a}$ controls the distribution of $Y(a)$. We assume that
$\theta$ is distributed over on an open ball of $\mathbb{R}^{k}$. 
\subsection{Asymptotic distribution based on fixed \texorpdfstring{$\theta$}{Lg}}
To start with, we treat $\theta$ as fixed parameters $\theta^{*}$,
which can be represented as known parameters, but may not be true
parameters. Extract the similar assumptions \citep{mack1979multivariate,abadie2016matching},
and impose some regularity conditions to derive the asymptotic distributions
of $\hat{\tau}_{\amw}^{\theta^{*}}$ and $\hat{\tau}_{\amw}^{t\theta^{*}}$
as follows.
\begin{assumption}\label{assumption2}
Set $\ensuremath{f(e_{i}^{*})}$ and $\ensuremath{\mathbb{E}\{R_{i}^{*}(A_{i})|\theta^{*}\}}$
are continually differential and bounded. Besides, $\ensuremath{R_{i}^{*}(A_{i})}$
and $\ensuremath{\mathbb{E}\{R_{i}^{*}(A_{i})^{3}|\theta^{*}\}}$
are uniformly bounded. Meanwhile, let $\ensuremath{u_{A_{i}}(X_{i},\beta_{A_{i}}^{*})}$ and $\ensuremath{\sigma_{A_{i}}^{2}(\beta_{A_{i}}^{*})}$
satisfy Lipschitz continuity conditions.
\end{assumption}
\citet{abadie2016matching} and \citet{yang2020multiply} demonstrated
these assumptions for PSM and double robust matching estimators, respectively.
They impose regularity conditions on moments and smoothness for inference.
We establish the theorems as follows. 
\begin{thm}
Under Assumptions \ref{assumption1}-\ref{assumption2}, as $n\rightarrow\infty$, if either the propensity
score model or the outcome model is correctly specified, let $\pi_{a}(e^{*})=P(a\mid e^{*})$, we have
\begin{equation}
n^{1/2}\left(\hat{\tau}_{\amw}^{\theta^{*}}-\tau\right)\rightarrow N\left(0,\Sigma_{\tau}^{\theta^{*}}\right),\label{eq:amwf_var}
\end{equation}
\[
\begin{aligned}\Sigma_{\tau}^{\theta^{*}} & =\text{\ensuremath{\mathbb{E}\left[\left\{  \frac{1}{\pi_{0}(e^{*})}\right\}  ^{2}\times\mathbb{{V}}\{(1-A)R^{*}(0)|X\}+\left\{  \frac{1}{\pi_{1}(e^{*})}\right\}  ^{2}\times\mathbb{{V}}\{AR^{*}(1)|X\}\right]}}\\
 & +\mathbb{E}[\mathbb{{V}}\{u_{1}(X,\beta_{1}^{*})-u_{0}(X,\beta_{0}^{*})-\tau\}].
\end{aligned}
\]
\end{thm}
We directly compare variances for the AMW, IPW, and
outcome regression estimators. In the case where the outcome model
is correctly specified, the efficiency of $\hat{\tau}_{\amw}^{\theta^{*}}$
may be lower than that of $\hat{\tau}_{\reg}^{\theta^{*}}$. Moreover,
if the outcome model is incorrectly specified, $\hat{\tau}_{\amw}^{\theta^{*}}$
maybe less efficient than the IPW estimator. However, the single-robust
outcome regression and IPW estimators are biased if the corresponding
parametric models are misspecified. In contrast, the linear expression
for AMW derived in the supplementary material is similar to AIPW,
indicating that AMW possesses the advantage of double robustness in
the large sample setting. Therefore, AMW consistently estimates the
ATE if either the propensity score or outcome model is correctly specified.
Specifically, if both nuisance models are correct, the variance of
the AMW estimator can be expressed as:
\[
\mathbb{E}\left[\frac{\mathbb{{V}}\{Y(0)\mid X\}}{\pi_{0}(e^{*})}+\frac{\mathbb{{V}}\{Y(1)|X\}}{\pi_{1}(e^{*})}+\{u_{1}(X,\beta_{1}^{*})-u_{0}(X,\beta_{0}^{*})-\tau\}^{2}\right].
\]
This variance achieves the semiparametric efficiency bound as the
variance of AIPW \citep{robins1994estimation,hahn1998role,tsiatis2007semiparametric}.
Hence, in the context of a large sample, the efficiency of the AMW estimator is comparable to that of the AIPW estimator.
\begin{thm}
Under Assumptions \ref{assumption1}-\ref{assumption2}, as $n_{1}\rightarrow\infty$, if either the
propensity score model or the outcome model is correctly specified, let $p=\mathbb{E}(A)$, we have
\begin{equation}
n^{1/2}\left(\hat{\tau}_{\amw}^{t,\theta^{*}}-\tau^{t}\right)\rightarrow N\left(0,\Sigma_{\tau}^{t,\theta^{*}}\right),\label{eq:amwtf_var}
\end{equation}
\begin{equation}
\begin{aligned}\Sigma_{\tau}^{t,\theta^{*}} & =\text{\ensuremath{\frac{1}{p^{2}}}\ensuremath{\mathbb{E}\left[\left\{  \frac{1-\pi_{0}(e^{*})}{\pi_{0}(e^{*})}\right\}  ^{2}\times\mathbb{{V}}\{(1-A)R^{*}(0)|X\}+\mathbb{{V}}\{AR^{*}(1)|X\}\right]}}\\
 & +\frac{1}{p^{2}}\mathbb{E}[\mathbb{{V}}\{YA-u_{0}\left(X,\beta_{0}^{*}\right)A-\tau^{t}\}].
\end{aligned}
\end{equation}
\end{thm}

\subsection{Asymptotic distribution based on estimated \texorpdfstring{$\theta$}{Lg}}

We acknowledge that $\theta$ is usually unknown and should be estimated
in practical applications. Therefore, this section will examine the
effect of estimating the nuisance parameters on the AMW estimators
within the framework of \citet{abadie2016matching} and  \citet{yang2020multiply}.
To investigate the limiting distributions of the AMW estimators with
estimated $\hat{\theta}$, we implement M-estimation to obtain $\hat{\theta}$.
The nuisance parameter $\hat{\theta}$ in candidate propensity score
and outcome models relies on the following estimating equations:
\[
\Psi({\theta})\equiv\frac{1}{n}\sum_{i=1}^{n}\psi(A_{i},X_{i},Y_{i};{\theta})=\frac{1}{n}\sum_{i=1}^{n}\left\{ \begin{array}{c}
\psi_{1}\left(A_{i},X_{i};\alpha\right)\\
\psi_{2}\left(A_{i},X_{i},Y_{i};\beta_{0}\right)\\
\psi_{3}\left(A_{i},X_{i},Y_{i};\beta_{1}\right)
\end{array}\right\} ,
\]
where
\begin{align*}
    \psi_{1}(A_{i},X_{i},Y_{i};\alpha)&=\frac{\partial e\left(X;\alpha\right)}{\partial\alpha}\frac{A-e\left(X,\alpha\right)}{e\left(X;\alpha\right)\left\{ 1-e(X,\alpha)\right\} }, \\
    \psi_{2}(A_{i},X_{i},Y_{i};\beta_{0})&=(1-A)\frac{\partial u_{0}\left(X;\beta_{0}\right)}{\partial\beta_{0}}\left\{ Y-u_{0}\left(X,\beta_{0}\right)\right\}, \\
    \psi_{3}(A_{i},X_{i},Y_{i};\beta_{1})&=A\frac{\partial u_{1}\left(X;\beta_{1}\right)}{\partial\beta_{1}}\left\{ Y-u_{1}\left(X,\beta_{1}\right)\right\}.
\end{align*}


Based on the Locally Asymptotically Normal (LAN) model, we define
the local parameter $\theta_{n}=\theta^{*}+h/\sqrt{n}$, where $\theta^{*}$
is the true parameter and $h$ is a constant. To invoke Le Cam's
third lemma \citep{le1990asymptotics,van2000asymptotic} for LAN model,
we first derive the limiting distribution between $n^{1/2}(\hat{\tau}_{\amw}^{\theta_{n}}-\tau^{\theta_{n}})$,
$n^{1/2}(\theta_{n}-\theta)$, and $\log(p^{\theta^{*}}/p^{\theta_{n}})$
under $p^{\theta_{n}}$, where $p^{\theta_{n}}$ is the probability
measure with $\theta_{n}$ and $p^{\theta^{*}}$ is the true probability measure of the random variables. Next, we obtain the limiting distribution
$n^{1/2}(\hat{\tau}_{\amw}^{\theta_{n}}-\tau)$ by Le Cam's third
lemma and derive the corresponding coefficients by utilizing martingale theory \citep{andreou2012alternative}. Lastly, the asymptotic distribution
for $n^{1/2}(\hat{\tau}_{\amw}^{\hat{\theta}}-\tau)$ can be approximated
by replacing $\theta_{n}$ with $\hat{\theta}$. 
\begin{thm}
Under Assumptions \ref{assumption1}-\ref{assumption2}, and regularity conditions specified in the
supplementary material, if either the propensity score model or the
outcome model is correctly specified, we have 
\begin{equation}
n^{1/2}\left(\hat{\tau}_{\amw}^{\hat{\theta}}-\tau\right)\rightarrow N\left(0,\Sigma_{\tau}^{\hat{\theta}}\right),\label{eq:atee_var}
\end{equation}
where $\Sigma_{\tau}^{\hat{\theta}}=\Sigma_{\tau}^{\theta^{*}}-C_{1}^{\top}I_{\theta^{*}}^{-1}C_{1}+C_{2}^{\top}\Sigma_{\theta^{*}}C_{2}$,
$I_{\theta^{*}}^{-1}=\mathbb{E}\left\{ \psi\left(A,X,Y;\theta^{*}\right)\psi\left(A,X,Y;\theta^{*}\right)^{\mathrm{\top}}\right\} ^{-1}$,
$\Sigma_{\theta^{*}}=\varphi_{\theta^{*}}^{-}I_{\theta^{*}}(\varphi_{\theta^{*}}^{-})^{\top}$
with $\varphi_{\theta^{*}}^{-}=\mathbb{E}\{\partial\psi(A,X,Y;\theta^{*})/\partial\theta\}^{-}$,
coefficients $C_{1}$ and $C_{2}$ are illustrated in the supplementary
material.
\end{thm}
The variance $\Sigma_{\tau}^{\hat{\theta}}$ incorporates two additional
terms, namely $-C_{1}^{\top}I_{\theta^{*}}^{-1}C_{1}$ and $C_{2}^{\top}\Sigma_{\theta^{*}}C_{2}$,
compared to $\Sigma_{\tau}^{\theta^{*}}$ in (\ref{eq:amwf_var}).
The term $-C_{1}^{\top}I_{\theta^{*}}^{-1}C_{1}$ is analogous to
the reduction term obtained by \citet{abadie2016matching}, which
demonstrates the correlation between the score function for $\theta$
in parametric models and the matching estimator, leading to a decrease
in estimated variance. Besides, the term $C_{2}^{\top}\Sigma_{\theta^{*}}C_{2}$
is similar to the variance inflation term obtained by \citet{yang2020multiply},
where $\tau$ relies on the nuisance parameters by equation $\mathbb{E}[u_{1}\left(X,\beta_{1}\right)-u_{0}\left(X,\beta_{0}\right)+AR/e(X,\alpha)-\allowbreak(1-A)R/\{1-e(X,\alpha)\}]$
with the misspecification of either the propensity score or outcome
model. The difference between $\Sigma_{\tau}^{\theta^{*}}$ in (\ref{eq:amwf_var})
and $\Sigma_{\tau}^{\hat{\theta}}$ in (\ref{eq:atee_var}) is unknown,
since the sum of the variance reduction and variance inflation terms
is unknown. Therefore, we cannot assert that the estimated score can
always improve estimation, and it might even reduce estimation efficiency.
\begin{thm}
Under Assumptions \ref{assumption1}-\ref{assumption2}, and regularity conditions specified in the
supplementary material, for $r\geq1,$ $0<n_{1}^{r}/n<\infty$, if
either the propensity score model or the outcome model is correctly
specified, we have 
\begin{equation}
n^{1/2}\left(\hat{\tau}_{\amw}^{t,\hat{\theta}}-\tau^{t}\right)\rightarrow N\left(0,\Sigma_{\tau}^{t,\hat{\theta}}\right),\label{eq:ateet_var}
\end{equation}
where \textup{$\Sigma_{\tau}^{t,\hat{\theta}}=\Sigma_{\tau}^{t,\theta^{*}}-C_{1}^{t,\top}I_{\theta^{*}}^{-1}C_{1}^{t}+C_{2}^{t,\top}\Sigma_{\theta^{*}}C_{2}^{t}$,}
$C_{1}^{t}$ and $C_{2}^{t}$ are illustrated in the supplementary
material.
\end{thm}
The extension of the discussion on large sample distribution properties
from the AMW estimator for ATE to that for ATT is straightforward.

\section{Simulation study} \label{sec:Simulation-study}

To examine the finite sample properties of the proposed AMW estimator
$\hat{\tau}_{\amw}$ for estimating the average treatment effect (ATE),
as well as other existing methods such as weighting and matching,
two experiments are conducted. One experiment uses the extreme propensity
score distribution, while the other uses the standard distribution.
The study aims to verify the AMW estimator's stable property, indicating
that it maintains a small mean squared error even when propensity
scores are close to zero or one. Additionally, the double robustness
property of the AMW estimator is to determine its consistency when
either the propensity score or the outcome model is correctly specified.
Lastly, the validity of the variance estimated by standard bootstrap
is also investigated. The simulations are conducted on 1000 Monte
Carlo simulated datasets for each scenario to obtain solid results.

\subsection{Simulation setting}

This simulation compares three types of estimators: 1) weighting type
estimators such as IPW and AIPW, 2) matching type estimator PSM, and
3) the AMW estimator with unfixed $K$ and the AMWF estimator, which
is the AMW estimator with fixed $K=1$. Parameter $K$ is selected
using cross-validation, repeated 25 times for robust bias estimation.

Set sample size to $n=1000$, and generate variables $Z_{j}\stackrel{\text{ }}{\stackrel{iid}{\sim}U[1-\sqrt{3},1+\allowbreak\sqrt{3}]}$,
where $j=1,..,12$. Transformations are applied to incorporate nonlinearity
and correlation into the confounder model: $X_{1}=\exp(Z_{1})$, $X_{2}=\exp(Z_{2})$,
$X_{3}=\log(Z_{3}+1)^{2}$, $X_{4}=\log(Z_{4}+1)^{2}$, $X_{5}=\sin(Z_{5}-Z_{6})$,
$X_{6}=\cos(Z_{5}+Z_{6})$, $X_{7}=\sin(Z_{7})$, $X_{8}=\cos(Z_{7}-1)$,
$X_{9}=(Z_{8}>0.4)$, $X_{10}=(Z_{8}>-0.4)$, $X_{11}=(Z_{9}>0.3)$,
$X_{12}=(Z_{10}>-0.3)$. Standardize the transformed $X$ by $\{X_{j}-\mathbb{E}(X_{j})\}/\sqrt{\{}\mathbb{V}(X_{j})\}$
in case some covariates may significantly influence the results. Potential
outcome models are generated as $Y(0)=\beta^{\top}X+\epsilon(0)$
and $Y(1)=\beta^{\top}X+\epsilon(1)$, where $\epsilon(0)\sim N(0,4)$,
$\epsilon(1)\sim N(0,1)$ and $\beta^{\top}=(1,1,1,1,-1,-1,-1,-1,1,-1,1,-1)\times0.2$.
Meanwhile, the probability of receiving treatment $A=1$ is $\textrm{logit}{}^{-1}(\alpha_{k}^{\top}X)$
with $\alpha_{k}^{\top}=(1,1,1,1,-1,-1,-1,-1,1,-1,1,-1)\times c_{k}$,
when $c_{1}=1$ induces a extreme propensity score distribution with
a heavy tail, and $c_{2}=0.3$ results in a standard propensity score
distribution with fewer extreme values.

Two model specifications are used to estimate the propensity scores:
1). a correctly specified logistic model $e(X,\alpha^{1})=\textrm{logit}{}^{-1}(\alpha^{1,\top}X)$;
2). a misspecified logistic model based on the original variables
$e(X,\alpha^{0})=\textrm{logit}{}^{-1}(\alpha^{0,\top}Z)$. Similarly,
we consider two model specifications for the outcomes: 1). a correctly
specified linear model $u_{a}(X,\beta_{a}^{1})=\beta_{a}^{1,\top}X$;
2). a misspecified linear model based on the original variables $u_{a}(X,\beta_{a}^{0})=\beta_{a}^{0,\top}Z$.
Therefore, we can use four combinations to fit the propensity score
and outcome models. To make it easier, each estimator's name is associated
with two numbers that indicate the choice of fitting models. The first
number indicates the selection for the propensity score model, while
the second number represents the choice for the outcome model. Here,
we use \textquotedbl 1\textquotedbl{} to denote the correctly specified
model and \textquotedbl 0\textquotedbl{} to denote the misspecified
model. For instance, \textquotedbl AMW01\textquotedbl{} represents
the AMW estimator that combines a misspecified propensity score model
and a correctly specified outcome model.

Standard nonparametric bootstrap approach is utilized to obtain the
variances of all estimators. Specifically, 500 bootstrap replicates
are conducted in the non-stable case with the correct propensity score
model to minimize the impact of extreme propensity scores. In other
cases, 100 bootstrap replicates are generated. Five summary statistics
are computed to evaluate the performance of the estimators:
\begin{itemize}
\item The sample mean value of 1000 simulated estimators (``mean''),
\item The sample standard deviation of 1000 simulated estimators (``sd''),
\item The average value of 1000 standard deviations from bootstrap replicates
(``bootsd''),
\item The MSE of 1000 simulated estimators (``mse''),
\item The coverage rates obtained by $95\%$ bootstrap quantiles (``cr'').
\end{itemize}

\subsection{Results}

\begin{figure}
\begin{centering}
\includegraphics[scale=0.88]{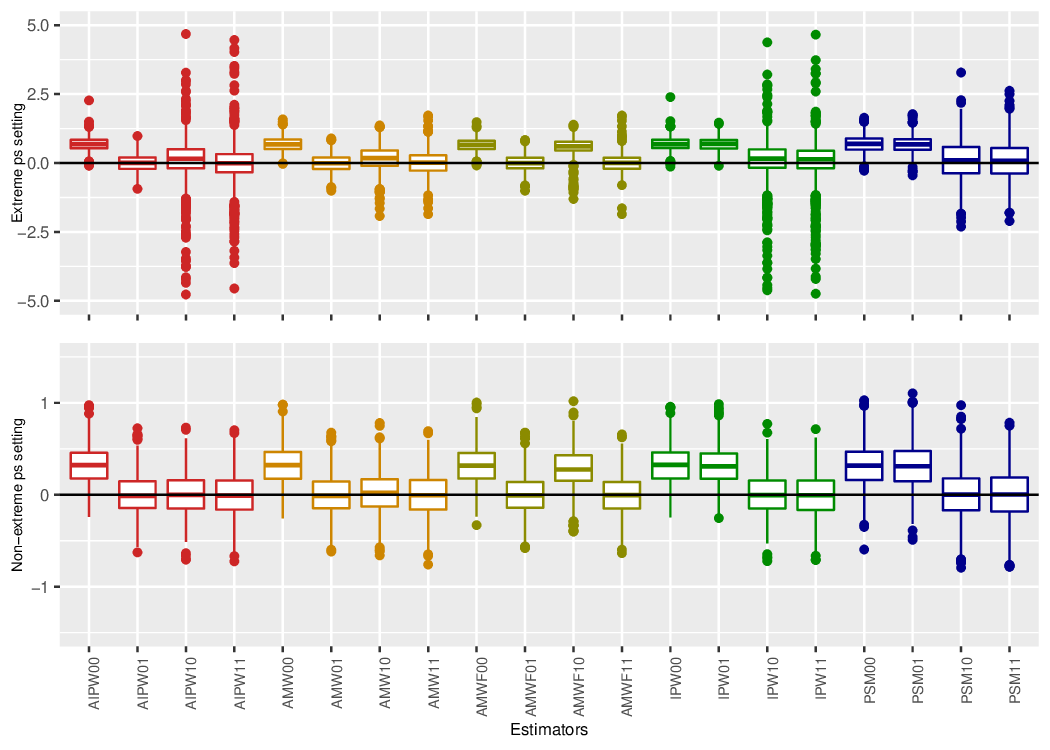}
\par\end{centering}
\caption{\label{fig:Box-plot}Box plot for ATE.}
\end{figure}

Table 1 summarizes the performance of different estimators based on
five summary statistics, while Figure \ref{fig:Box-plot} displays
the box plots for each estimator. The results indicate that the AMW
estimators consistently achieve the minimum MSE or near-minimum MSE
among all the estimators. When the propensity score distribution is
not extreme and the population is adequate, the difference between
the AMW estimator and the locally efficient AIPW estimator is negligible.
However, when the propensity scores are extreme, the AMW estimators
outperform the AIPW estimator, as the inverse of the extreme weights
in the weighting type estimators leads to unstable estimates. We remove
the outliers beyond the range of $(-5,5)$ on the y-axis in Figure
\ref{fig:Box-plot} to provide a more harmonized presentation. The
PSM estimator consistently has a larger MSE than the AMW estimators,
which justifies the efficiency gain from the augmented terms in the
AMW estimators.

The results obtained from both scenarios indicate that the AMW-type
estimators are doubly robust, which is a significant advantage over
singly robust estimators like the IPW and PSM estimators. Additionally,
we observe that the bias of the AMWF10 estimator is not negligible
in both settings, which supports the use of unfixed $K$ in the AMW
estimators. Specifically, the AMWF10 estimator can be considered as
the PSM estimator for $K=1$ with an incorrect outcome term, which
introduces a new bias for AMWF.

Furthermore, we find that the standard deviation estimated by bootstrap
for the AMWF estimator is consistently higher than the simulated standard
deviation. In contrast, the smoothed AMW estimator with unfixed $K$
can provide a more reliable and consistent standard deviation estimation
from the standard bootstrap. As a result, the AMW estimator with unfixed
$K$ shows a more promising coverage rate compared to other matching
type estimators.

\section{Real data analysis}\label{sec:Real-data-analysis}

In this section, we utilize AMW estimators to study the causal effects
in three publicly accessible datasets \citep{loh2021confounder}.
To generate the AMW estimators, we first posit propensity score and
outcome models. Although the outcome model selections can vary across
studies, logistic regression is used to estimate all propensity scores.
Additionally, all covariates are scaled before being incorporated
into the models to prevent certain variables from dominating the results.
Once the AMW estimators are obtained, we apply standard bootstrap
with 500 repetitions to investigate the standard errors. Since researchers cannot verify the
actual causal effects, we assess
the matching performances using standardized differences of covariates
\citep{abadie2011bias} between the treatment and control groups.
Reduced differences after matching indicate a successful balance between
the two groups. This paper includes plots of the absolute standardized
differences, with blue dots representing differences before matching
and red dots representing differences after matching for corresponding
covariates. Tables displaying standardized differences are presented
in the supplementary material.

\subsection{Labor training program}

\cite{lalonde1986evaluating} and \cite{dehejia1999causal} analyzed
the National Supported Work (NSW-DW) dataset, which is a subset of
a labor training program. The dataset, available in the \textquotedbl Matching\textquotedbl{}
package in R, aims to examine the causal effect of a job training
program. There are a total of 445 individuals in the NSW-DW dataset,
with 185 individuals in the treatment group. The outcome is represented
by the actual earnings of each person in 1978, and seven covariates
are taken into account in the linear outcome model. Specifically,
the second-order terms of two numeric covariates, namely age and actual
earnings in 1975, are included in both the propensity score and outcome
models \citep{dehejia1999causal}.

\begin{figure}
\centering{}\includegraphics[scale=0.88]{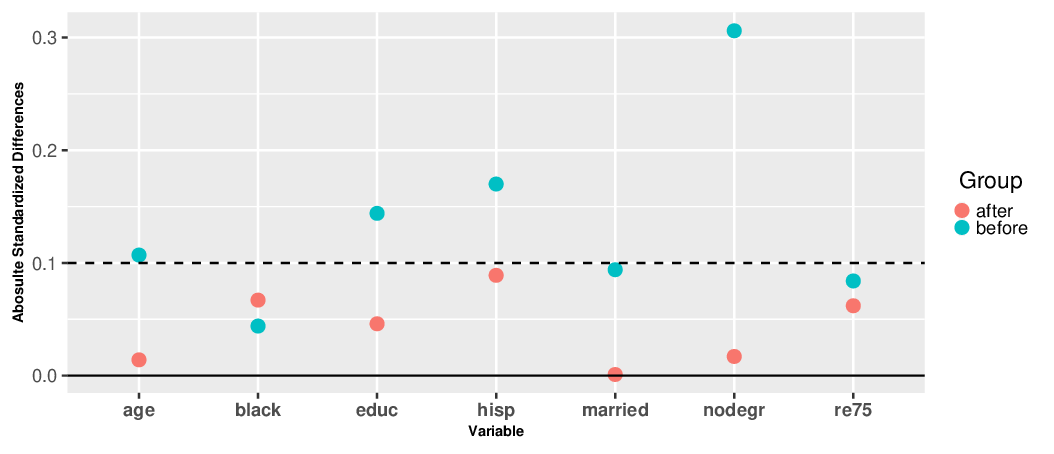}
\caption{Standardized differences plot for Labor Training Program data.}
\label{fig:ll}
\end{figure}

To provide a more comprehensive evaluation of this program, we analyze
the ATT for participating in the job training program, as the ATE
may not be sufficient. Figure \ref{fig:ll} displays the standardized
differences, illustrating that the AMW estimator achieves greater
balance of covariates between the two groups. The ATT, which measures
the average earning increase for the treated in one year, is 1872.53
dollars, with a standard error of 692.70. The empirical $95\%$ confidence
interval, which reflects the uncertainty surrounding the ATT, is $(514.84,3230.22)$.
These results suggest that the job training program has a significant
impact on the average earning of the treated group, consistent with
the NSW-DW dataset's estimated ATT of 1794.34 \citep{otsu2016bootstrap}.

\subsection{AIDS clinical trials group study }

\cite{hammer1996trial} designed a double-blind study to assess the
efficacy of AIDS treatments, comparing the effects of a single treatment
of either zidovudine or didanosine to the combination treatment of
zidovudine plus didanosine or zalcitabine. The dataset used for this
study, AIDS Clinical Trials Group Study 175 (ACTG175), is available
in the R package \textquotedbl speff2trial.\textquotedbl{} As HIV
can attack the immune system and reduce CD4 cell counts, the primary
outcome of interest in our design is the difference between CD4 cell
counts after 96 weeks and the baseline. Units with missing outcomes
are excluded to maintain the validity of the modeling process, resulting
in a reduced model with 1342 individuals. While the original study
was designed as a randomized experiment, missing values posed challenges
when comparing the two therapy groups directly. Thus, it is necessary
to employ the AMW estimator to eliminate confounding bias and obtain
reliable effects. The outcome model utilizes a linear model with all
14 covariates. 

\begin{figure}
\centering{}\includegraphics[scale=0.7]{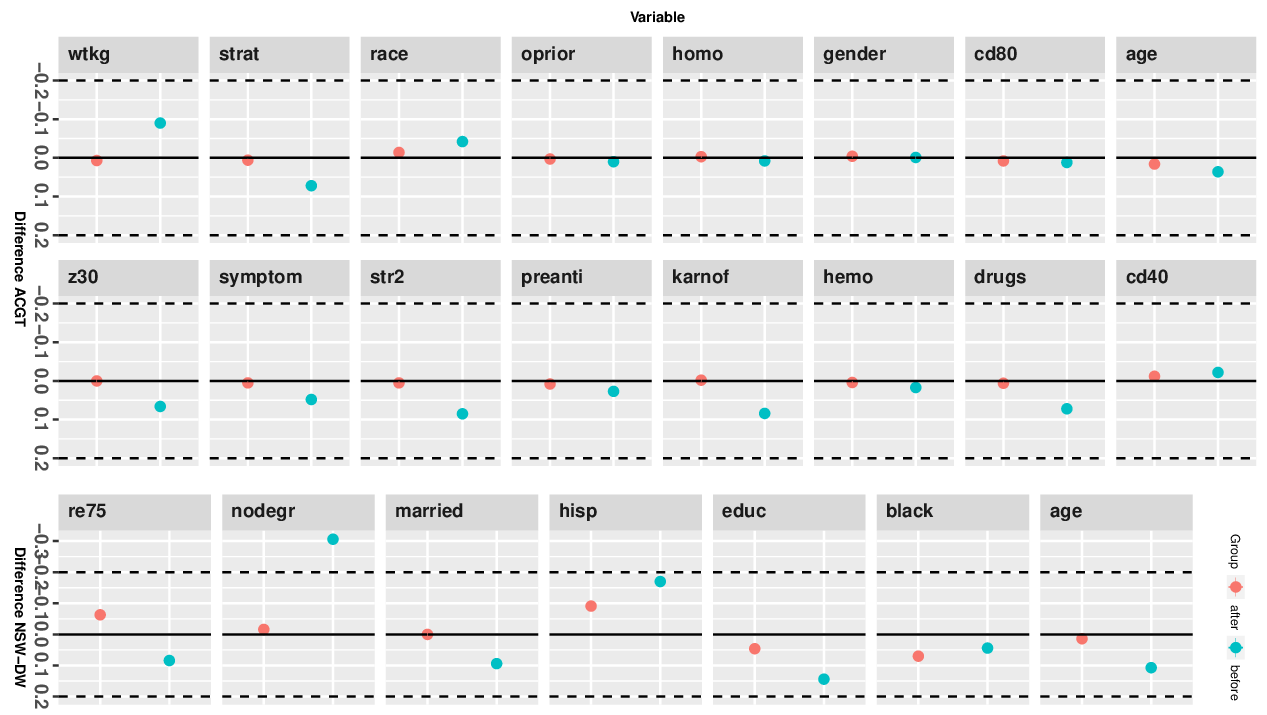}\caption{Standardized differences plot for ACTG data.}
\label{fig:acgt}
\end{figure}

Figure \ref{fig:acgt} displays the standardized differences for covariates
before and after matching, which demonstrates that the AMW estimator
effectively enhances the balance between the two groups. The average
treatment effect (ATE) is calculated to be 40.852 with a standard
error of 8.220, yielding a $95\%$ empirical lower bound for ATE of
24.740. This result, which is greater than zero, suggests a significant
improvement in CD4 cell counts resulting from combination therapy.

\subsection{Right heart catheterization}

\cite{connors1996effectiveness} and \cite{hirano2001estimation}
analyzed the RHC dataset to investigate the safety and effectiveness
of the Right Heart Catheterization procedure for severely ill patients
in ICUs. The RHC procedure was previously believed to increase patient
survival rates, but subsequent reanalysis by \cite{hirano2001estimation}
suggested otherwise. We aim to use the AMW estimator to analyze the
RHC dataset and evaluate its effectiveness on this challenging dataset.
The RHC dataset, which contains 5735 units with 63 variables, is available
at https://hbiostat.org/data/. For our analysis, we select 50 variables
and convert categorical variables into dummy variables for both the
propensity score and outcome models. We model the outcome variable,
which is a binary indicator of whether the patient survives for 30
days (the survival patient denoted as ``1'', and the dead patient
denoted as ``0''), using logistic regression. Our goal is to estimate
the causal risk difference by comparing the survival probabilities
between the two experimental groups.

\begin{figure}
\begin{centering}
\includegraphics[angle=360,scale=0.55]{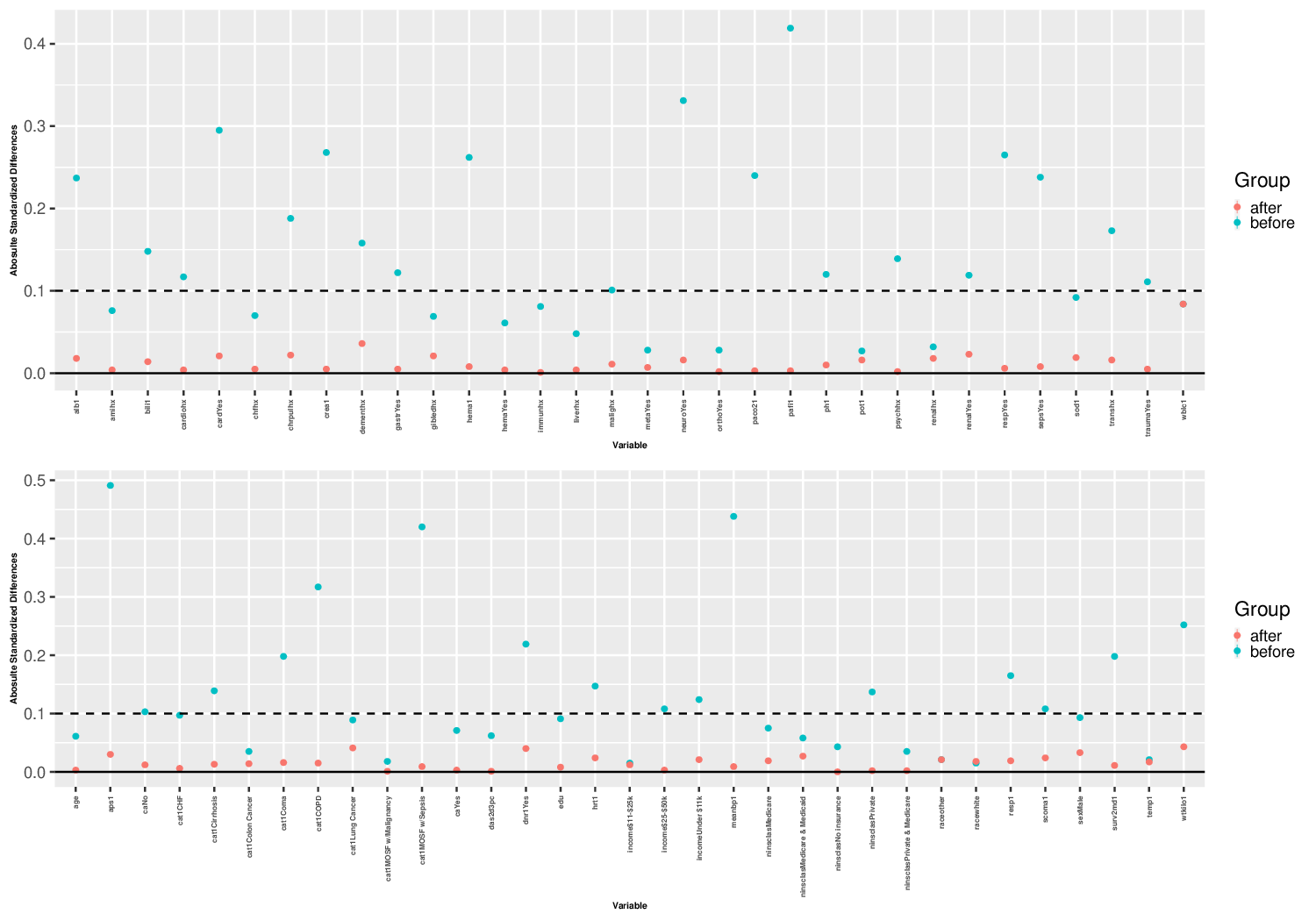}
\par\end{centering}
\caption{Standardized differences plot for RHC data.}
\label{fig:rhc}
\end{figure}

Figure \ref{fig:rhc} demonstrates that the use of the AMW estimator
results in reduced differences between the observed covariates in
the two groups before and after matching. Our estimated ATT of -0.0641
is consistent with the effects obtained by overlap and optimal matching
estimators, as reported by \cite{li2016balancing}. However, the
standard error for our estimator, 0.0241, is higher than those obtained
by the other two estimators. Our findings suggest that the RHC procedure
is associated with increased risk for patients, which contrasts with
the effects estimated by traditional approaches that do not account
for confounding bias.

\section{Discussion}\label{sec:Discussion}
We propose a new approach called Augmented Match Weighted (AMW) estimator
for estimating general causal estimands. In comparison with PSM, AMW
achieves the semi-parametric efficiency bound for unfixed $K$ and
exhibits double robustness due to the inclusion of additional augmentation
terms. Unlike the AIPW approach, AMW avoids the need to directly invert
the propensity scores, thereby increasing its stability. Our simulation
results demonstrate that the naive bootstrap method can be used to
obtain variance estimates for AMW. Overall, the AMW estimator is a
practical and powerful alternative to existing methods for causal
inference. However, to address the issue of computationally tractable problems, we typically fix the value of $K$ to estimate the variance through bootstrap in CV. While this approach does not pose any significant trouble in practice, a more standard method is to allow $K$ to vary independently in each bootstrap iteration to achieve a more precise selection.

Assumption \ref{assumption1} requires all confounders should be measured
and included to construct the propensity score and outcome models
so that people may include all variables in models to guarantee the
assumption hold. However, \cite{brookhart2006variable} emphasized
that instrumental variables may not improve the estimation of treatment effects for the propensity score model since they are uncorrelated with outcomes.
The variables related to outcomes can provide better estimation even
under model misspecifications. Hence, \cite{zhang2021practical}
recommended a simple yet effective strategy that employs lasso \cite{hastie2009elements}
for variable selection in the outcome model, followed by using the
chosen features with appropriately tuned propensity scores to improve
the efficiency of matching. This approach can significantly reduce
bias and variance in PSM. Likewise, one can investigate how various
variables might affect AMW, where AMW could be less susceptible to
instrumental variables. Multiple variable selection techniques, such
as lasso and Adaptive lasso, can be combined with AMW to enhance its
efficiency.

Meanwhile, \cite{chernozhukov2018double} investigated the application
of debiased/double machine learning (DML) methods for estimating high-dimensional
nuisance parameters, and established broad theoretical foundations
with weak assumptions. This led to the integration of the DML framework
into AMW, allowing for the use of modern nonparametric techniques
(such as xgboost and random forest) in estimating causal effects.

On the other hand, \cite{yang2016propensity} proposed a framework
that employs PSM to estimate treatment effects in scenarios with more
than two treatment options, and AMW could be similarly adapted to
handle such cases. Additionally, it is interesting to extend the AMW estimator to longitudinal \cite{burvzkova2007longitudinal} or survival analysis
\cite{choi2017estimating}.

\section{The Article Header Information}

\subsection{Acknowledgements} 
T.X. was supported by N.S.F. DEB-1754142. S.Y. was partially supported by NIH 1R01AG066883
and 1R01ES031651. 

\bibliographystyle{tfcad}
\bibliography{Ch2}

@book{ding2024first,
  title={A first course in causal inference},
  author={Ding, Peng},
  year={2024},
  publisher={Chapman and Hall/CRC}
}

@article{lin2023estimation,
  title={Estimation based on nearest neighbor matching: from density ratio to average treatment effect},
  author={Lin, Zhexiao and Ding, Peng and Han, Fang},
  journal={Econometrica},
  volume={91},
  number={6},
  pages={2187--2217},
  year={2023},
  publisher={Wiley Online Library}
}

@Article{abadie2006large,
  Title                    = {Large sample properties of matching estimators for average treatment effects},
  Author                   = {Abadie, Alberto and Imbens, Guido W},
  Journal                  = {Econometrica},
  Year                     = {2006},
  Pages                    = {235--267},
  Volume                   = {74},
  Publisher                = {Wiley Online Library}
}

@Article{abadie2008failure,
  Title                    = {On the failure of the bootstrap for matching estimators},
  Author                   = {Abadie, Alberto and Imbens, Guido W},
  Journal                  = {Econometrica},
  Year                     = {2008},
  Pages                    = {1537--1557},
  Volume                   = {76},
  Publisher                = {Wiley Online Library}
}

@Article{abadie2011bias,
  Title                    = {Bias-corrected matching estimators for average treatment effects},
  Author                   = {Abadie, Alberto and Imbens, Guido W},
  Journal                  = {Journal of Business \& Economic Statistics},
  Year                     = {2011},
  Pages                    = {1--11},
  Volume                   = {29},
  Publisher                = {Taylor \& Francis}
}

@Article{abadie2012martingale,
  author    = {Abadie, Alberto and Imbens, Guido W},
  title     = {A martingale representation for matching estimators},
  pages     = {833--843},
  volume    = {107},
  journal   = {J Am Stat Assoc},
  publisher = {Taylor \& Francis},
  year      = {2012},
}

@Article{abadie2016matching,
  Title                    = {Matching on the estimated propensity score},
  Author                   = {Abadie, Alberto and Imbens, Guido W},
  Journal                  = {Econometrica},
  Year                     = {2016},
  Pages                    = {781--807},
  Volume                   = {84},
  Publisher                = {Wiley Online Library}
}

@Article{andreou2012alternative,
  author    = {Andreou, Elena and Werker, Bas JM},
  title     = {An alternative asymptotic analysis of residual-based statistics},
  pages     = {88--99},
  volume    = {94},
  journal   = {Rev Econ Stat},
  publisher = {MIT Press},
  year      = {2012},
}

@Article{bang2005doubly,
  Title                    = {Doubly robust estimation in missing data and causal inference models},
  Author                   = {Bang, Heejung and Robins, James M},
  Journal                  = {Biometrics},
  Year                     = {2005},
  Pages                    = {962--973},
  Volume                   = {61},
  Publisher                = {Wiley Online Library}
}

@Article{brookhart2006variable,
  author    = {Brookhart, M Alan and Schneeweiss, Sebastian and Rothman, Kenneth J and Glynn, Robert J and Avorn, Jerry and St{\"u}rmer, Til},
  title     = {Variable selection for propensity score models},
  pages     = {1149--1156},
  volume    = {163},
  journal   = {Am J Epidemiol},
  publisher = {Oxford University Press},
  year      = {2006},
}

@Article{burvzkova2007longitudinal,
  Title                    = {Longitudinal data analysis for generalized linear models with follow-up dependent on outcome-related variables},
  Author                   = {Buzkova, Petra and Lumley, Thomas},
  Journal                  = {Canadian Journal of Statistics},
  Year                     = {2007},
  Pages                    = {485--500},
  Volume                   = {35},
  Publisher                = {Wiley Online Library}
}

@Article{cao2009improving,
  Title                    = {Improving efficiency and robustness of the doubly robust estimator for a population mean with incomplete data},
  Author                   = {Cao, Weihua and Tsiatis, Anastasios A and Davidian, Marie},
  Journal                  = {Biometrika},
  Year                     = {2009},
  Pages                    = {723--734},
  Volume                   = {96},
  Publisher                = {Biometrika Trust}
}

@Article{chernozhukov2018double,
  Title                    = {Double/debiased machine learning for treatment and structural parameters},
  Author                   = {Chernozhukov, Victor and Chetverikov, Denis and Demirer, Mert and Duflo, Esther and Hansen, Christian and Newey, Whitney and Robins, James},
  Journal                  = {The Econometrics Journal},
  Year                     = {2018},
  Pages                    = {1--68},
  Volume                   = {21},
  Publisher                = {Wiley Online Library}
}

@Article{choi2017estimating,
  Title                    = {Estimating the causal effect of treatment in observational studies with survival time end points and unmeasured confounding},
  Author                   = {Choi, Jaeun and O'Malley, A James},
  Journal                  = {Journal of the Royal Statistical Society: Series C},
  Year                     = {2017},
  Pages                    = {159--185},
  Volume                   = {66},
  Publisher                = {Wiley Online Library}
}

@Article{dehejia&wahba98,
  author  = {Dehejia, R.~H. and Wahba, S.},
  title   = {Propensity score matching methods for non-experimental causal studies},
  pages   = {151-161},
  volume  = {84},
  journal = {Rev Econ Stat},
  year    = {2002},
}

@Article{efron1979bootstrap,
  Title                    = {Bootstrap methods: another look at the jackknife},
  Author                   = {Efron, Bradley},
  Journal                  = {Annals of Statistics},
  Year                     = {1979},
  Pages                    = {1--26},
  Volume                   = {7},
  Booktitle                = {Breakthroughs in Statistics},
  Publisher                = {Springer}
}

@Article{hahn1998role,
  Title                    = {On the role of the propensity score in efficient semiparametric estimation of average treatment effects},
  Author                   = {Hahn, Jinyong},
  Journal                  = {Econometrica},
  Year                     = {1998},
  Pages                    = {315--331},
  Volume                   = {66},
  Publisher                = {JSTOR}
}

@Article{han2013estimation,
  Title                    = {Estimation with missing data: beyond double robustness},
  Author                   = {Han, Peisong and Wang, Lu},
  Journal                  = {Biometrika},
  Year                     = {2013},
  Pages                    = {417--430},
  Volume                   = {100},
  Publisher                = {Biometrika Trust}
}

@Book{hastie2009elements,
  Title                    = {The Elements of Statistical Learning},
  Author                   = {Hastie, Trevor and Tibshirani, Robert and Friedman, Jerome and Hastie, T and Friedman, J and Tibshirani, R},
  Publisher                = {Springer},
  Year                     = {2009},
  Volume                   = {2}
}

@Article{heckman1997matching,
  Title                    = {Matching as an econometric evaluation estimator: Evidence from evaluating a job training programme},
  Author                   = {Heckman, James J and Ichimura, Hidehiko and Todd, Petra E},
  Journal                  = {The Review of Economic Studies},
  Year                     = {1997},
  Pages                    = {605--654},
  Volume                   = {64},
  Publisher                = {Oxford University Press}
}

@Article{hirano2001estimation,
  Title                    = {Estimation of causal effects using propensity score weighting: An application to data on right heart catheterization},
  Author                   = {Hirano, Keisuke and Imbens, Guido W},
  Journal                  = {Health Services and Outcomes Research Methodology},
  Year                     = {2001},
  Pages                    = {259--278},
  Volume                   = {2},
  Publisher                = {Springer}
}

@Book{imbens2015causal,
  Title                    = {{Causal Inference in Statistics, Social, and Biomedical Sciences}},
  Author                   = {Imbens, Guido W and Rubin, Donald B},
  Publisher                = {Cambridge University Press},
  Year                     = {2015},
  Address                  = {Cambridge UK}
}

@Article{kang2007demystifying,
  Title                    = {Demystifying double robustness: A comparison of alternative strategies for estimating a population mean from incomplete data},
  Author                   = {Kang, Joseph DY and Schafer, Joseph L},
  Journal                  = {Statistical Science},
  Year                     = {2007},
  Pages                    = {523--539},
  Volume                   = {22},
  Publisher                = {JSTOR}
}

@Article{lalonde1986evaluating,
  Title                    = {Evaluating the econometric evaluations of training programs with experimental data},
  Author                   = {LaLonde, Robert J},
  Journal                  = {The American Economic Review},
  Year                     = {1986},
  Pages                    = {604--620},
  Volume                   = {76},
  Publisher                = {JSTOR}
}

@Book{le1990asymptotics,
  Title                    = {Asymptotics in Statistics: Some Basic Concepts},
  Author                   = {Le Cam, Lucien and Yang, Grace Lo},
  Publisher                = {Springer: Berlin},
  Year                     = {1990}
}

@Article{leacy2014joint,
  author    = {Leacy, Finbarr P and Stuart, Elizabeth A},
  title     = {On the joint use of propensity and prognostic scores in estimation of the average treatment effect on the treated: a simulation study},
  pages     = {3488--3508},
  volume    = {33},
  journal   = {Stat Med},
  publisher = {Wiley Online Library},
  year      = {2014},
}

@Article{li2016balancing,
  author    = {Li, Fan and Morgan, Kari Lock and Zaslavsky, Alan M},
  title     = {Balancing covariates via propensity score weighting},
  pages     = {doi:10.1080/01621459.2016.1260466},
  journal   = {J Am Stat Assoc},
  publisher = {Taylor \& Francis},
  year      = {2017},
}

@Article{mack1979multivariate,
  Title                    = {Multivariate k-nearest neighbor density estimates},
  Author                   = {Mack, YP and Rosenblatt, Murray},
  Journal                  = {Journal of Multivariate Analysis},
  Year                     = {1979},
  Pages                    = {1--15},
  Volume                   = {9},
  Publisher                = {Elsevier}
}

@Article{otsu2016bootstrap,
  author    = {Otsu, Taisuke and Rai, Yoshiyasu},
  title     = {Bootstrap inference of matching estimators for average treatment effects},
  pages     = {1720--1732},
  volume    = {112},
  journal   = {J Am Stat Assoc},
  publisher = {Taylor \& Francis},
  year      = {2017},
}

@Article{robins1994estimation,
  author    = {Robins, James M and Rotnitzky, Andrea and Zhao, Lue Ping},
  title     = {Estimation of regression coefficients when some regressors are not always observed},
  pages     = {846--866},
  volume    = {89},
  journal   = {J Am Stat Assoc},
  publisher = {Taylor \& Francis Group},
  year      = {1994},
}

@Article{rosenbaum1983central,
  Title                    = {The central role of the propensity score in observational studies for causal effects},
  Author                   = {Rosenbaum, Paul R and Rubin, Donald B},
  Journal                  = {Biometrika},
  Year                     = {1983},
  Pages                    = {41--55},
  Volume                   = {70},
  Publisher                = {Biometrika Trust}
}

@Article{rubin1973matching,
  Title                    = {Matching to remove bias in observational studies},
  Author                   = {Rubin, Donald B},
  Journal                  = {Biometrics},
  Year                     = {1973},
  Pages                    = {159--183},
  Volume                   = {29},
  Publisher                = {JSTOR}
}

@Book{rubin2006matched,
  Title                    = {{Matched Sampling for Causal Effects}},
  Author                   = {Rubin, Donald B},
  Publisher                = {Cambridge University Press},
  Year                     = {2006},
  Address                  = {Cambridge, England}
}

@Article{stuart2010matching,
  Title                    = {Matching methods for causal inference: A review and a look forward},
  Author                   = {Stuart, Elizabeth A},
  Journal                  = {Statistical Science},
  Year                     = {2010},
  Pages                    = {1--21},
  Volume                   = {25},
  Publisher                = {NIH Public Access}
}

@Book{tsiatis2007semiparametric,
  Title                    = {{Semiparametric Theory and Missing Data}},
  Author                   = {Tsiatis, Anastasios},
  Publisher                = {Springer, New York},
  Year                     = {2006}
}

@Book{van2000asymptotic,
  Title                    = {Asymptotic Statistics},
  Author                   = {van der Vaart},
  Publisher                = {Cambridge university press},
  Year                     = {2000},
  Address                  = {Cambridge: Cambridge University Press},
  Volume                   = {3}
}

@Article{yang2016propensity,
  Title                    = {Propensity score matching and subclassification in observational studies with multi-level treatments},
  Author                   = {Yang, Shu and Imbens, Guido W and Cui, Zhanglin and Faries, Douglas E and Kadziola, Zbigniew},
  Journal                  = {Biometrics},
  Year                     = {2016},
  Pages                    = {1055--1065},
  Volume                   = {72},
  Publisher                = {Wiley Online Library}
}

@Article{yangkim2016,
  Title                    = {A semiparametric inference to regression analysis with missing covariates in survey data},
  Author                   = {Yang, Shu and Kim, Jae Kwang},
  Journal                  = {Statistica Sinica},
  Year                     = {2017},
  Note                     = {Accepted for publication},
  Pages                    = {261--285},
  Volume                   = {27}
}

@Book{guo2014propensity,
  author    = {Guo, Shenyang and Fraser, Mark W},
  title     = {Propensity Score Analysis: Statistical Methods and Applications},
  publisher = {Thousand Oaks, CA: SAGE},
  volume    = {11},
  year      = {2014},
}

@Article{frolich2004finite,
  author    = {Fr{\"o}lich, Markus},
  title     = {Finite-sample properties of propensity-score matching and weighting estimators},
  pages     = {77--90},
  volume    = {86},
  journal   = {Rev Econ Stat},
  publisher = {MIT Press},
  year      = {2004},
}

@Article{dehejia1999causal,
  author    = {Dehejia, Rajeev H and Wahba, Sadek},
  title     = {Causal effects in nonexperimental studies: Reevaluating the evaluation of training programs},
  pages     = {1053--1062},
  volume    = {94},
  journal   = {J Am Stat Assoc},
  publisher = {Taylor \& Francis Group},
  year      = {1999},
}

@article{yang2020multiply,
  title={Multiply robust matching estimators of average and quantile treatment effects},
  author={Yang, Shu and Zhang, Yunshu},
  journal={arXiv preprint arXiv:2001.06049},
  year={2020}
}

@article{ju2019adaptive,
  title={On adaptive propensity score truncation in causal inference},
  author={Ju, Cheng and Schwab, Joshua and van der Laan, Mark J},
  journal={Statistical methods in medical research},
  volume={28},
  number={6},
  pages={1741--1760},
  year={2019},
  publisher={SAGE Publications Sage UK: London, England}
}

@article{hammer1996trial,
  title={A trial comparing nucleoside monotherapy with combination therapy in HIV-infected adults with CD4 cell counts from 200 to 500 per cubic millimeter},
  author={Hammer, Scott M and Katzenstein, David A and Hughes, Michael D and Gundacker, Holly and Schooley, Robert T and Haubrich, Richard H and Henry, W Keith and Lederman, Michael M and Phair, John P and Niu, Manette and others},
  journal={New England Journal of Medicine},
  volume={335},
  number={15},
  pages={1081--1090},
  year={1996},
  publisher={Mass Medical Soc}
}

@article{connors1996effectiveness,
  title={The effectiveness of right heart catheterization in the initial care of critically III patients},
  author={Connors, Alfred F and Speroff, Theodore and Dawson, Neal V and Thomas, Charles and Harrell, Frank E and Wagner, Douglas and Desbiens, Norman and Goldman, Lee and Wu, Albert W and Califf, Robert M and others},
  journal={Jama},
  volume={276},
  number={11},
  pages={889--897},
  year={1996},
  publisher={American Medical Association}
}

@article{cochran1973controlling,
  title={Controlling bias in observational studies: A review},
  author={Cochran, William G and Rubin, Donald B},
  journal={Sankhy{\=a}: The Indian Journal of Statistics, Series A},
  pages={417--446},
  year={1973},
  publisher={JSTOR}
}

@article{loh2021confounder,
  title={Confounder selection strategies targeting stable treatment effect estimators},
  author={Loh, Wen Wei and Vansteelandt, Stijn},
  journal={Statistics in Medicine},
  volume={40},
  number={3},
  pages={607--630},
  year={2021},
  publisher={Wiley Online Library}
}

@article{zhang2021practical,
  title={Practical recommendations on double score matching for estimating causal effects},
  author={Zhang, Yunshu and Yang, Shu and Ye, Wenyu and Faries, Douglas E and Lipkovich, Ilya and Kadziola, Zbigniew},
  journal={Statistics in medicine},
  year={2021},
  publisher={Wiley Online Library}
}

@article{rothenhausler2020model,
  title={Model selection for estimation of causal parameters},
  author={Rothenh{\"a}usler, Dominik},
  journal={arXiv preprint arXiv:2008.12892},
  year={2020}
}

@article{cui2019selective,
  title={Selective machine learning of doubly robust functionals},
  author={Cui, Yifan and Tchetgen, Eric Tchetgen},
  journal={arXiv preprint arXiv:1911.02029},
  year={2019}
}

@article{brookhart2006semiparametric,
  title={A semiparametric model selection criterion with applications to the marginal structural model},
  author={Brookhart, M Alan and Van Der Laan, Mark J},
  journal={Computational statistics \& data analysis},
  volume={50},
  number={2},
  pages={475--498},
  year={2006},
  publisher={Elsevier}
}

@article{rolling2014model,
  title={Model selection for estimating treatment effects},
  author={Rolling, Craig A and Yang, Yuhong},
  journal={Journal of the Royal Statistical Society: Series B (Statistical Methodology)},
  volume={76},
  number={4},
  pages={749--769},
  year={2014},
  publisher={Wiley Online Library}
}

@article{newey2018cross,
  title={Cross-fitting and fast remainder rates for semiparametric estimation},
  author={Newey, Whitney K and Robins, James R},
  journal={arXiv preprint arXiv:1801.09138},
  year={2018}
}

\appendix

\begin{table}[H] 
\caption{Simulation results.} 
\resizebox{\columnwidth}{!}{%
\begin{tabular}{|l|lllll|lllll|}
\hline Name   & \multicolumn{5}{l|}{Extreme propensity score distribution}    & \multicolumn{5}{l|}{Non-extreme propensity score distribution}                              \\ \hline type   & mean   & sd    & bootsd & mse   & cr    & mean   & sd    & bootsd & mse   & \multicolumn{1}{l|}{cr}    \\ AMW11  & 0.01   & 0.427 & 0.401  & 0.182 & 0.946 & -0.002 & 0.23  & 0.231  & 0.053 & \multicolumn{1}{l|}{0.938} \\ IPW11  & -0.044 & 1.248 & 0.768  & 1.558 & 0.846 & -0.008 & 0.233 & 0.222  & 0.055 & \multicolumn{1}{l|}{0.927} \\ AIPW11 & -0.051 & 1.129 & 0.704  & 1.277 & 0.93  & -0.006 & 0.229 & 0.219  & 0.053 & \multicolumn{1}{l|}{0.919} \\ PSM11  & 0.087  & 0.7   & 0.598  & 0.497 & 0.94  & 0.005  & 0.271 & 0.257  & 0.073 & \multicolumn{1}{l|}{0.967} \\ AMWF11 & 0.003  & 0.319 & 0.6    & 0.101 & 0.952 & -0.004 & 0.214 & 0.256  & 0.046 & \multicolumn{1}{l|}{0.963} \\ \hline AMW01  & -0.009 & 0.302 & 0.309  & 0.091 & 0.95  & 0      & 0.22  & 0.224  & 0.048 & \multicolumn{1}{l|}{0.943} \\ IPW01  & 0.682  & 0.231 & 0.235  & 0.518 & 0.17  & 0.314  & 0.2   & 0.2    & 0.139 & \multicolumn{1}{l|}{0.634} \\ AIPW01 & -0.001 & 0.293 & 0.296  & 0.086 & 0.938 & -0.001 & 0.214 & 0.21   & 0.046 & \multicolumn{1}{l|}{0.915} \\ PSM01  & 0.674  & 0.3   & 0.296  & 0.544 & 0.315 & 0.315  & 0.234 & 0.236  & 0.154 & \multicolumn{1}{l|}{0.719} \\ AMWF01 & -0.006 & 0.279 & 0.339  & 0.078 & 0.956 & 0.002  & 0.212 & 0.245  & 0.045 & \multicolumn{1}{l|}{0.96}  \\ \hline AMW10  & 0.156  & 0.432 & 0.392  & 0.21  & 0.926 & 0.023  & 0.224 & 0.229  & 0.051 & \multicolumn{1}{l|}{0.957} \\ IPW10  & 0.071  & 1.388 & 0.732  & 1.929 & 0.871 & 0      & 0.224 & 0.222  & 0.05  & \multicolumn{1}{l|}{0.943} \\ AIPW10 & 0.085  & 1.364 & 0.735  & 1.867 & 0.891 & 0.005  & 0.223 & 0.221  & 0.05  & \multicolumn{1}{l|}{0.948} \\ PSM10  & 0.103  & 0.705 & 0.602  & 0.507 & 0.961 & 0.004  & 0.265 & 0.257  & 0.07  & \multicolumn{1}{l|}{0.973} \\ AMWF10 & 0.591  & 0.289 & 0.609  & 0.433 & 0.973 & 0.28   & 0.209 & 0.259  & 0.122 & \multicolumn{1}{l|}{0.975} \\ \hline AMW00  & 0.682  & 0.248 & 0.26   & 0.527 & 0.232 & 0.32   & 0.203 & 0.214  & 0.144 & \multicolumn{1}{l|}{0.66}  \\ IPW00  & 0.692  & 0.239 & 0.236  & 0.536 & 0.155 & 0.321  & 0.199 & 0.199  & 0.143 & \multicolumn{1}{l|}{0.602} \\ AIPW00 & 0.691  & 0.244 & 0.241  & 0.537 & 0.163 & 0.322  & 0.199 & 0.199  & 0.143 & \multicolumn{1}{l|}{0.609} \\ PSM00  & 0.693  & 0.305 & 0.299  & 0.574 & 0.307 & 0.317  & 0.234 & 0.237  & 0.155 & \multicolumn{1}{l|}{0.72}  \\ AMWF00 & 0.66   & 0.22  & 0.3    & 0.484 & 0.325 & 0.318  & 0.198 & 0.235  & 0.14  & \multicolumn{1}{l|}{0.724} \\ \hline 
\end{tabular} 
}

\label{Summary}
\end{table}

\end{document}


\articletype{ORIGINAL ARTICLES}

\title{Supplementary Materials for ''Augmented Match Weighted Estimators: New Methods for Estimating Average Treatment Effects Under Extreme Propensity Scores''}


\maketitle

\section{Proof for Lemma 2}

We divide AMW estimator into two part as $B(K)$ and $C$,
where 
$$B(K)=\frac{1}{n} \sum_{i=1}^{n}\left\{\hat{m}_{a_{i}}(e^{*}_{i})A_{i}+\hat{m}_{a_{i}}(e^{*}_{i})(1-A_{i})\right\}.$$

\subsection {Bias and Variance}
Since both $R^{*}_{i}(a_{i})$ and $e^{*}_{i}$ are given, we can apply the theorems from Mack to illustrate the bias of $B(K)$ as:
$$
\begin{aligned}
24f^{3}(e^{*}) \times Bias\{B(K)\}=\left\{ (m_{1}f)''(e^{*})-m_{1}(e^{*})f''(e^{*})\right\} *\frac{K^{2}}{n_{1}^{2}}\\
-\left\{ (m_{0}f)''(e^{*})-m_{0}(e^{*})f''(e^{*})\right\} *\frac{K^{2}}{n_{0}^{2}}.
\end{aligned}
$$
$$n \times Var\left\{ B(K)\right\} =\frac{\mathbb{{V}}(AR^{*}(1)|e^{*})+\mathbb{{V}}\{(1-A)R^{*}(0)|e^{*}\}}{K}
$$

\section{Proof for Theorem 1}
\subsection{Rewrite the estimator with KNN}
For simplicity, $\hat{m}_{A_{i}}(e^{*}_{i})$ is denoted as:
$$\hat{m}_{A_{i}}(e^{*}_{i})=\frac{\sum_{j \in \mathcal{J}_{K}(i)} M_{R_{e_{i}}}\left({e}^{*}_{i}-{e}^{*}_{j}\right) {{R}}^{*}_{j}(A_{i})I_{(A_{j}=A_{i})}}{\sum_{j \in \mathcal{J}_{K}(i)} M_{R_{e_{i}}}\left({e}^{*}_{i}-{e}^{*}_{j}\right)I_{(A_{j}=A_{i})}}=\frac{h_{n}({e}^{*}_{i})}{f_{n}({e}^{*}_{i})}.$$
\\
Let $\hat{\tau}^{\theta^{*}}_{\mathrm{AMW}}$ be the estimator with given score,and rewrite our estimator into three parts:
$$\hat{\tau}^{\theta^{*}}_{\mathrm{AMW}}=\hat{\tau}^{\theta^{*}}_{\mathrm{reg}}+\hat{\tau}^{\theta^{*}}_{\mathrm{1}}-\hat{\tau}^{\theta^{*}}_{\mathrm{0}},$$
\\
where 
$$\hat{\tau}^{\theta^{*}}_{\mathrm{1}}=\frac{1}{n} \sum_{i=1}^{n}\left\{R^{*}_{i}(1)A_{i}+\hat{m}_{A_{i}}(e^{*}_{i})(1-A_{i})\right\},$$

$$\hat{\tau}^{\theta^{*}}_{\mathrm{0}}=\frac{1}{n} \sum_{i=1}^{n}\left\{\hat{m}_{A_{i}}(e^{*}_{i})A_{i}+R_{i}^{*}(0)(1-A_{i})\right\}.$$
Let's  make expectation that $m_{A_{i}}(e^{*}_{i})=\mathbb{E}\left\{R^{*}_{i}(A_{i})|e^{*}_{i}\right\}$, and rewrite 

$$\hat{\tau}^{\theta^{*}}_{\mathrm{1}}=\frac{1}{n} \sum_{i=1}^{n}\left[R_{i}^{*}(1)A_{i}+m_{1}(e^{*}_{i})(1-A_{i})+\left\{\hat{m}_{1}(e^{*}_{i})-m_{1}(e^{*}_{i})\right\}(1-A_{i})\right],$$

$$\hat{\tau}^{\theta^{*}}_{\mathrm{0}}=\frac{1}{n} \sum_{i=1}^{n}\left[R^{*}_{i}(0)(1-A_{i})+m_{0}(e^{*}_{i})A_{i}+\left\{\hat{m}_{0}(e^{*}_{i})-m_{0}(e^{*}_{i})\right\}A_{i}\right].$$
\\
Detect the relationship between $\hat{m}_{1}(e^{*}_{i})-m_{1}(e^{*}_{i})$ by lemma from Mack(1981):
$$h_{n}(\mathbf{x})=f(\mathbf{x})\pi(\mathbf{x}) m(\mathbf{x})+O\left(\frac{K}{n}\right)^{2}+O\left(\frac{1}{K}\right),$$
$$f_{n}(\mathbf{x})=f(\mathbf{x})\pi(\mathbf{x}) +O\left(\frac{K}{n}\right)^{2}+O\left(\frac{1}{K}\right).$$
\\
Denote $\pi_{0}(e^{*}_{i})=\mathbb{P}\left\{1_{(A_{i}=0)} \mid e^{*}_{i}\right\}$, $\pi_{1}(e^{*}_{i})=\mathbb{P}\left\{1_{(A_{i}=1)}\mid e^{*}_{i}\right\}$. Hence by Taylor expansion:
\\
$$\hat{m}_{1}(e^{*}_{i})-m_{1}(e^{*}_{i})=\frac{1}{f({e}^{*}_{i})\pi_{1}(e^{*}_{i})}
\left\{h_{n}({e}^{*}_{i})-f_{n}({e}^{*}_{i}) m_{1}({e}^{*}_{i})\right\}+O\left(\frac{K}{n}\right)^{2}+O\left(\frac{1}{K}\right).$$\\
\\
We express the similar form for $\hat{m}_{0}(e^{*}_{i})-m_{0}(e^{*}_{i})$.

\subsection{U stat}
Basing on the definition, we know $m_{Ai}(e^{*}_{i})=\mathbb{E}\left\{R^{*}_{i}(A_{i})|e^{*}_{i}\right\}=0$,
$$
\begin{aligned}
\hat{\tau}^{\theta^{*}}_{\mathrm{AMW}}&=\frac{1}{n}\sum_{i=1}^{n}\left\{u_{1}(X_{i},{\beta}^{*}_{1})-u_{0}(X_{i},{\beta}^{*}_{0})
+R^{*}_{i}(1)A_{i}-R^{*}_{i}(0)(1-A_{i})\right\}\\
&-\sum_{i=1}^{n}(A_{i}\frac{1}{n}\frac{1}{f(e^{*}_{i})\pi_{0}(e^{*}_{i})} \sum_{j=1}^{n}\left[1_{(A_{j}=0)}  M_{R_{e^{*}_{i} }}\left(e^{*}_{i}-e^{*}_{j}\right)\left\{R^{*}_{j}(0)-\hat{m}_{0}(e^{*}_{i})\right\}\right])\\
&+\sum_{i=1}^{n}((1-A_{i})\frac{1}{n}\frac{1}{f(e^{*}_{i})\pi_{1}(e^{*}_{i})} \sum_{j=1}^{n}\left[ 1_{(A_{j}=1)}  M_{R_{e^{*}_{i} }}\left(e^{*}_{i}-e^{*}_{j}\right)\left\{R^{*}_{j}(1)-\hat{m}_{1}(e^{*}_{i})\right\}\right] )\\
&+O\left(\frac{K}{n}\right)^{2}+O\left(\frac{1}{K}\right).
\end{aligned}
$$
\\
Define
$$U_{1}=\frac{1}{n^2}\sum_{i=1}^{n}A_{i} \frac{1}{f(e^{*}_{i})}{\pi_{0}(e^{*}_{i})} \sum_{j=1}^{n}  1_{(A_{j}=0)} M_{R_{e^{*}_{i} }}\left\{f(e^{*}_{i})-e^{*}_{j}\right\}\left\{R^{*}_{j}(0)-\hat{m}_{0}(e^{*}_{i})\right\}.$$
\\
Using U statistics(van der Vaart):
$$
\begin{aligned}
U_{1}=\frac{1}{n(n-1)} \sum_{i=1}^{n}\sum_{j \neq i} h\left({Z}_{i}, {Z}_{j}\right),
\end{aligned}
$$
where $ {Z}_{i}=(R^{*}_{i}(0),A_{i},e^{*}_{i})$, and 
$$
\begin{aligned}
h\left({Z}_{i}, {Z}_{j}\right)&=\frac{1}{2}[A_{i}1_{(A_{j}=0)}  \frac{1}{f\left(e^{*}_{i}\right)\pi_{0}(e^{*}_{i})}M_{R_{e^{*}_{i} }}\left(e^{*}_{i}-e^{*}_{j}\right)\left\{R^{*}_{j}(0)-\hat{m}_{0}(e^{*}_{i})\right\}\\
&+A_{j} 1_{(A_{i}=0)} \frac{1}{f\left(e^{*}_{j}\right)\pi_{0}(e^{*}_{j})}M_{R_{e^{*}_{i} }}\left(e^{*}_{j}-e^{*}_{i}\right)\left\{R^{*}_{i}(0)-\hat{m}_{0}(e^{*}_{j})\right\}]\\
&=\frac{1}{2}\left(\zeta_{i j}+\zeta_{j i}\right).
\end{aligned}
$$

$$
\begin{aligned}
\mathbb{E}(\zeta_{i j}|{Z}_{i})=&\mathbb{E}\left(A_{i} \frac{1}{f\left(e^{*}_{i}\right)\pi_{0}(e^{*}_{i})}\mathbb{E}\left[1_{(A_{j}=0)}M_{R_{e^{*}_{i} }}\left(e^{*}_{i}-e^{*}_{j}\right)\left\{R^{*}_{j}(0)-\hat{m}_{0}(e^{*}_{i})\right\}|Z_{i}\right]\right)\\
=&\mathbb{E}\left[A_{i} \frac{1}{f\left(e^{*}_{i}\right)\pi_{0}(e^{*}_{i})}f\left(e^{*}_{j}\right)\pi_{0}(e^{*}_{j})\left\{m_{0}(e^{*}_{i})-\hat{m}_{0}(e^{*}_{i})\right\}\right]\\
=&O\left(\frac{K}{n}\right)^{2}+O\left(\frac{1}{K}\right).
\end{aligned}
$$

$$
\begin{aligned}
\mathbb{E}(\zeta_{j i}|{Z}_{i})=&\mathbb{E}\left[A_{j} \frac{1}{f\left(e^{*}_{j}\right)\pi_{0}(e^{*}_{j})}1_{(A_{i}=0)}M_{R_{e^{*}_{j} }}\left(e^{*}_{j}-e^{*}_{i}\right)\left\{R^{*}_{i}(0)-\hat{m}_{0}(e^{*}_{i})\right\}|Z_{i}\right]\\
=&(1-A_{i})\mathbb{E}\left(\mathbb{E}\left[A_{j} \frac{1}{f\left(e^{*}_{j}\right)\pi_{0}(e^{*}_{j})}M_{R_{e^{*}_{j} }}\left(e^{*}_{j}-e^{*}_{i}\right)\left\{R^{*}_{i}(0)-\hat{m}_{0}(e^{*}_{i})\right\}|Z_{i},R_{e^{*}_{j}}\right]|Z_{i}\right)\\
=&(1-A_{i}) \frac{1-\pi_{0}(e^{*}_{i})}{\pi_{0}(e^{*}_{i})} \left\{R^{*}_{i}(0)-\hat{m}_{0}(e^{*}_{i})\right\}+O\left(\frac{K}{n}\right)^{2}+O\left(\frac{1}{K}\right).
\end{aligned}
$$
Hence,
$$
\begin{aligned}
\hat{\tau}^{\theta^{*}}_{\mathrm{AMW}}-\tau&=\frac{1}{n}\sum_{i=1}^{n}\left\{Y_{i}A_{i}-Y_{i}(1-A_{i})-u_{0}(X_{i},{\beta}^{*}_{0})A_{i}+u_{1}(X_{i},{\beta}^{*}_{1})(1-A_{i})\right\}-\tau
\\
&-\frac{1}{n} \sum_{i=1}^{n}(1-A_{i}) \frac{1-\pi_{0}(e^{*}_{i})}{\pi_{0}(e^{*}_{i})}R^{*}_{i}(0)+\frac{1}{n} \sum_{i=1}^{n}A_{i} \frac{1-\pi_{1}(e^{*}_{i})}{\pi_{1}(e^{*}_{i})}R^{*}_{i}(1)\\
&+O\left(\frac{K}{n}\right)^{2}+O\left(\frac{1}{K}\right).
\end{aligned}
$$

$$
\begin{aligned}
\hat{\tau}^{\theta^{*}}_{\mathrm{AMW}}-\tau&=\frac{1}{n}\sum_{i=1}^{n}\left\{Y_{i}A_{i}-Y_{i}(1-A_{i})-u_{0}(X_{i},{\beta}^{*}_{0})A_{i}+u_{1}(X_{i},{\beta}^{*}_{1})(1-A_{i})\right\}-\tau \\
&+\frac{1}{n} \sum_{i=1}^{n}(1-A_{i})R^{*}_{i}(0)-\frac{1}{n}\sum_{i=1}^{n}A_{i}R^{*}_{i}(1)\\
&-\frac{1}{n} \sum_{i=1}^{n}(1-A_{i}) \frac{1}{\pi_{0}(e^{*}_{i})}R^{*}_{i}(0)+\frac{1}{n} \sum_{i=1}^{n}A_{i} \frac{1}{\pi_{1}(e^{*}_{i})}R^{*}_{i}(1)\\
&+O\left(\frac{K}{n}\right)^{2}+O\left(\frac{1}{K}\right).
\end{aligned}
$$

$$
\begin{aligned}
\hat{\tau}^{\theta^{*}}_{\mathrm{AMW}}-\tau&=\frac{1}{n} \sum_{i=1}^{n}\left\{u_{1}\left(X_{i}, {\beta}^{*}_{1}\right)-u_{0}\left(X_{i}, {\beta}^{*}_{0}\right)\right\}-\tau
\\
&-\frac{1}{n} \sum_{i=1}^{n}(1-A_{i}) \frac{1}{\pi_{0}(e^{*}_{i})}R^{*}_{i}(0)+\frac{1}{n} \sum_{i=1}^{n}A_{i} \frac{1}{\pi_{1}(e^{*}_{i})}R^{*}_{i}(1)\\
&+O\left(\frac{K}{n}\right)^{2}+O\left(\frac{1}{K}\right).
\end{aligned}
$$
This linear form is identical to that of AIPW estimators. Hence, the AMW estimator shares a similar property with AIPW estimators.

\subsection{The double robust property}
The decomposition for the linear form is obtained as $\tau_{1}^{*}$ and $\tau_{0}^{*}$, where 

$$\tau_{1}^{*}=\frac{1}{n}\sum_{i=1}^{n}\left[u_{1}\left(X_{i},{\beta}_{1}^{*}\right)+A\frac{1}{\pi_{1}(e_{i}^{*})}\{Y(1)-u_{1}\left(X_{i},{\beta}_{1}^{*}\right)\}\right]  ,$$
$$\tau_{0}^{*}=\frac{1}{n}\sum_{i=1}^{n}-\left[u_{0}\left(X_{i},{\beta}_{0}^{*}\right)+(1-A_{i})\frac{1}{\pi_{0}(e_{i}^{*})}\{Y(0)-u_{0}\left(X_{i},{\beta}_{0}^{*}\right)\right]. $$
And the expectations of $\tau_{1}^{*}$ and $\tau_{0}^{*}$ are:
$$\mathbb{E}(\tau_{1}^{*})=\mathbb{E}\{Y(1)\}+\mathbb{E}\left[\frac{\{A-\pi_{1}(e^{*})\}}{\pi_{1}(e^{*})}\{Y(1)-u_{1}(X,\beta_{1}^{*})\}\right],$$
$$-\mathbb{E}(\tau_{0}^{*})=\mathbb{E}\{Y(0)\}+\mathbb{E}\left[\frac{\{1-A-\pi_{0}(e^{*})\}}{\pi_{0}(e^{*})}\{Y(0)-u_{0}(X,\beta_{0}^{*})\}\right].$$

If only the outcome model is correct, we obtain $\mathbb{{E}}\{Y(1)|X\}=u_{1,true}(X,\beta_{1})$, so
$$\mathbb{{E}}\left(\mathbb{E}\left[\frac{\{A-\pi_{1}(e^{*})\}}{\pi_{1}(e^{*})}\{Y(1)-u_{1}(X,\beta_{1}^{*})\}|X\right]\right)=0.$$

We also obtain $\mathbb{{E}}\{Y(0)|X\}=u_{0,true}(X,\beta_{0})$, so
$$\mathbb{{E}}\left(\mathbb{E}\left[\frac{\{1-A-\pi_{0}(e^{*})\}}{\pi_{0}(e^{*})}\{Y(0)-u_{0}(X,\beta_{0}^{*})\}|X\right]\right)=0.$$

Therefore, the AMW estimator is unbiased when only the outcome model is correct.

If only the propensity score model is correct, we obtain $$\mathbb{{E}}\left(\mathbb{E}\left[\frac{\{A-\pi_{1}(e^{*})\}}{\pi_{1}(e^{*})}|X\right]\right)=\allowbreak\mathbb{E}\left\{ \frac{\pi_{1,true}(e^{*})-\pi_{1}(e^{*})}{\pi_{1}(e^{*})}\right\},$$
$$\mathbb{{E}}\left(\mathbb{E}\left[\frac{\{1-A-\pi_{0}(e^{*})\}}{\pi_{0}(e^{*})}|X\right]\right)=\allowbreak\mathbb{E}\left\{ \frac{\pi_{0,true}(e^{*})-\pi_{0}(e^{*})}{\pi_{0}(e^{*})}\right\} ,$$
so
$$\mathbb{{E}}\left(\mathbb{E}\left[\frac{\{A-\pi_{1}(e^{*})\}}{\pi_{1}(e^{*})}\{Y(1)-u_{1}(X,\beta_{1}^{*})\}|X\right]\right)=0,$$
$$\mathbb{{E}}\left(\mathbb{E}\left[\frac{\{1-A-\pi_{0}(e^{*})\}}{\pi_{0}(e^{*})}\{Y(0)-u_{0}(X,\beta_{0}^{*})\}|X\right]\right)=0.$$

Therefore, the AMW estimator is unbiased when only the propensity score model is correct.

The variance for the linear form  is:
$$
\begin{aligned}
\Sigma_{\tau}^{\theta^{*}}	&=\mathbb{E}\left[\left\{\frac{1}{\pi_{0}(e^{*})}\right\}^{2}\mathbb{{V}}\{(1-A)R^{*}(0)\mid X\}+\left\{\frac{1}{\pi_{1}(e^{*})}\right\}^{2}\mathbb{{V}}\{AR^{*}(1)|X\}\right]\\
	&+\mathbb{E}\left[\mathbb{{V}}\{u_{1}(X,\beta_{1}^{*})-u_{0}(X,\beta_{0}^{*})-\tau\}\right].
\end{aligned}
$$

\section{Proof for Theorem 2}
The idea is identical to the proof for Theorem 1, and we keep the central step for obtaining Theorem 2.

\subsection{Rewrite the estimator with KNN}
Let $\hat{\tau}^{t,\theta*}_{\mathrm{AMW}}$ be the estimator with given score, and rewrite our estimator into three parts:
$$\hat{\tau}^{t,\theta^{*}}_{\mathrm{AMW}}=\hat{\tau}^{t,\theta^{*}}_{\mathrm{reg}}+\hat{\tau}^{t,\theta^{*}}_{\mathrm{1}}-\hat{\tau}^{t,\theta^{*}}_{\mathrm{0}},$$
\\
where 
$$\hat{\tau}^{t,\theta^{*}}_{\mathrm{1}}=\frac{1}{n_{1}} \sum_{i=1}^{n}R^{*}_{i}(1)A_{i},$$

$$\hat{\tau}^{t,\theta^{*}}_{\mathrm{0}}=\frac{1}{n_{1}} \sum_{i=1}^{n}\hat{m}_{A_{i}}(e^{*}_{i})A_{i}.$$

\subsection {KNN transform}
Let's  make expectation that $m_{A_{i}}(e_{i}^{*})=\mathbb{E}(R^{*}_{i}(A_{i})|e^{*}_{i})$, and rewrite 

$$\hat{\tau}^{t,\theta^{*}}_{\mathrm{0}}=\frac{1}{n_{1}} \sum_{i=1}^{n}\left[m_{0}(e^{*}_{i})A_{i}+\left\{\hat{m}_{0}(e^{*}_{i})-m_{0}(e^{*}_{i})\right\}A_{i}\right].$$
\\
Set $\pi_{0}(e^{*}_{i})=\mathbb{P}\left\{1_{(A_{i}=0)} \mid e^{*}_{i}\right\}$, hence by Taylor expansion:
\\
$$\hat{m}_{0}(e^{*}_{i})-m_{0}(e^{*}_{i})=\frac{1}{f(e^{*}_{i})\pi_{0}(e^{*}_{i})}
\left\{h_{n}(e^{*}_{i})-f_{n}(e^{*}_{i}) m_{0}(e^{*}_{i})\right\}+O\left(\frac{K}{n_{1}}\right)^{2}+O\left(\frac{1}{K}\right).$$\\

\subsection{U stat}
We rewrite the estimator:
$$
\begin{aligned}
\hat{\tau}^{t,\theta^{*}}_{\mathrm{AMW}}
&= \frac{1}{n_{1}}\sum_{i=1}^{n}\left\{A_{i}Y_{i}-A_{i}u_{0}(X_{i},\beta^{*}_{0})+A_{i}R^{*}_{i}(1)-m_{0}(e^{*}_{i})A_{i}\right\} \\
&\quad - \frac{1}{n_{1}}\sum_{i=1}^{n}\left[A_{i}\frac{1}{f(e^{*}_{i})\pi_{0}(e^{*}_{i})}\sum_{j=1}^{n}\left\{1_{(A_{j}=0)} M_{R_{e^{*}_{i}}}(e^{*}_{i}-e^{*}_{j})\left(R^{*}_{j}(0)-\hat{m}_{0}(e^{*}_{i})\right)\right\}\right]\\
&\quad + O\left(\frac{K}{n_{1}}\right)^{2}+O\left(\frac{1}{K}\right).
\end{aligned}
$$

Using U statistics (van der Vaart):
$$
\begin{aligned}
\mathbb{E}(\zeta_{j i}\mid Z_{i})
&=\mathbb{E}\left[A_{j} \frac{1}{f\left(e^{*}_{j}\right)\pi_{0}(e^{*}_{j})}1_{(A_{i}=0)}M_{R_{e^{*}_{j}}}\left(e^{*}_{j}-e^{*}_{i}\right)\left\{R^{*}_{i}(0)-\hat{m}_{0}(e^{*}_{i})\right\}\mid Z_{i}\right]\\
&=(1 - A_{i})\,\mathbb{E}\left(\mathbb{E}\left[A_{j} \frac{1}{f\left(e^{*}_{j}\right)\pi_{0}(e^{*}_{j})}M_{R_{e^{*}_{j}}}\left(e^{*}_{j}-e^{*}_{i}\right)\left\{R^{*}_{i}(0)-\hat{m}_{0}(e^{*}_{i})\right\}\mid Z_{i}, R_{e^{*}_{j}}\right]\mid Z_{i}\right)\\
&=(1 - A_{i})\,\frac{1 - \pi_{0}(e^{*}_{i})}{\pi_{0}(e^{*}_{i})}\left\{R^{*}_{i}(0)-\hat{m}_{0}(e^{*}_{i})\right\}+O\left(\frac{K}{n_{1}}\right)^{2}+O\left(\frac{1}{K}\right).
\end{aligned}
$$

$$
\begin{aligned}
\hat{\tau}^{t,\theta^{*}}_{\mathrm{AMW}}-\tau^{t}&=\frac{1}{n_{1}}\sum_{{i}=1}^{n}\left\{A_{i}Y_{i}-A_{i}u_{0}(X_{i,},\beta^{*}_{0})+A_{i}R^{*}_{i}(1)-m_{0}(e^{*}_{i})A_{i}\right\}-\tau^{t}
\\
&-\frac{1}{n_{1}} \sum_{i=1}^{n}(1-A_{i}) \frac{1-\pi_{0}(e^{*}_{i})}{\pi_{0}(e^{*}_{i})}\left\{R_{i}(0)-\hat{m}_{0}(e^{*}_{i})\right\}\\
&+O\left(\frac{K}{n_{1}}\right)^{2}+O\left(\frac{1}{K}\right).
\end{aligned}
$$
\\

$$
\begin{aligned}
\hat{\tau}^{t,\theta^{*}}_{\mathrm{AMW}}-\tau^{t}&=\frac{1}{n_{1}}\sum_{{i}=1}^{n}\left\{A_{i}Y_{i}-A_{i}u_{0}(X_{i,},\beta^{*}_{0})+A_{i}R^{*}_{i}(1)\right\}-\tau^{t}
\\
&-\frac{1}{n_{1}} \sum_{i=1}^{n}(1-A_{i}) \frac{1-\pi_{0}(e^{*}_{i})}{\pi_{0}(e^{*}_{i})}R^{*}_{i}(0)\\
&+O\left(\frac{K}{n_{1}}\right)^{2}+O\left(\frac{1}{K}\right).
\end{aligned}
$$

\section{Proof for theorem 3}
Suppose parameters are estimated, we can abstract the basic frame from Abadie(2016) and van der Vaart(2000) to do inference with $\theta^{\top}=(\alpha^{\top},\beta_{0}^{\top},\beta_{1}^{\top})$.  Under the true models, define $\theta^{*\top}=(\alpha^{*\top},\beta_{0}^{*\top},\beta_{1}^{*\top})$ satisfy $\mathbb{E}\left\{\psi \left(A, X, Y ; \theta^{*}\right)\right\}=0$, which can be obtain by M-estimation:
$$
\Psi(\theta^{*})\equiv \mathbb{E}\left\{\psi \left(A, X, Y ; \theta^{*}\right)\right\}=\mathbb{E}\left\{\begin{array}{c}
\psi _{1}\left(A, X , \alpha^{*}\right) \\
\psi _{2}\left(A, X, Y , \beta_{0}^{*}\right) \\
\psi _{3}\left(A, X, Y , \beta_{1}^{*}\right)
\end{array}\right\}=0,
$$

\noindent {\bf [Assumption 6] } 
1. Estimator $\hat{\theta}$ is a zero of $\Psi_{n}$ converge in probability to $\theta^{*}$  a zero of $\Psi$; 2. $\varphi(\theta)$ is score function as $E(\partial \psi (A,X,Y;\theta) /  \partial \theta)$, which is nonsingular around $\theta^{*}$.\\
\\
By Taylor expansion
$$\sqrt{n}(\hat{\theta}-\theta^{*})=-\varphi_{\theta^{*}}^{-}*\Psi(\theta^{*}) +o_{p}(1),$$
and $$\Psi(\theta^{*}) \overset{d}{\rightarrow} N(0,I_{\theta^{*}}),$$
apply delta method, as $n\rightarrow \infty$,
$$\sqrt{n}(\hat{\theta}-\theta^{*}) \overset{d}{\rightarrow} N(0,\varphi_{\theta^{*}}^{-}I_{\theta^{*}}(\varphi_{\theta^{*}}^{-})^{\top}),$$
where $I_{\theta^{*}}$ is  $\mathbb{E}(\psi(\theta^{*})\psi(\theta^{*})^{\top})$, and denote 
$\varphi_{\theta^{*}}^{-}I_{\theta^{*}}(\varphi_{\theta^{*}}^{-})^{\top}=\Sigma_{\theta^{*}}$\\

\noindent {\bf [Assumption 7] Regularity Conditions for Local Normality } 
Define any $\theta_{n}$ around fixed $\theta^{*}$, where $\theta_{n}=\theta^{*}+h/ \sqrt{n}$,  which can be treated as local shift from $\theta^{*}$.\\
1. $S^{\theta_{n}}_{i}$ are scores basing $\theta_{n}$, which have a compact and convex supporting set and are 
continuous and bounded away from zero. 2. $u(X_{i},\theta_{n})$ and $\sigma^{2}(\theta_{n})$ are satisfying Lipschitz continuity conditions. 3. $E(R(A_{i})^{3}|\theta_{n})$ is uniformly bounded. \\
\\
Gain the likelihood ratio based on $\theta^{*}$ and  $\theta_{n}$. Let $p_{\theta^{*}}$ be a density from $\mathbb{P}^{n}_{\theta^{*}}$, and take $l_{\theta^{*}}=log(p_{\theta^{*}}).$
Apply the  asymptotic local normal theory from Van der Vaart(2000) and log density from Yang(2020):
$$log\left(\frac{\mathbb{p}_{\theta^{*}}}{\mathbb{p}_{\theta_{n}}}\right)=- h^{T}\varphi(\theta^{*})*I_{\theta^{*}}^{-}*\Psi(\theta^{*})  -\frac{1}{2} h^{T} \Sigma_{\theta^{*}} h+o_{p}(1).$$

\subsection{Asymptotic joint distribution}
Under distribution of $\theta_{n}$, the joint distribution can be derived as:
$$
\left\{\begin{array}{c}
n^{1 / 2}\left(\hat{\tau}_{\mathrm{AMW}}^{\theta_{n}}-\tau^{\theta_{n}}\right) \\
n^{1 / 2}\left(\hat{\theta}-\theta_{n}\right) \\
\log \left( {p}^{\theta^{*}} /  {p}^{\theta_{n}}\right)
\end{array}\right\}\overset{d}{\rightarrow} \mathcal{N}\left\{\left(\begin{array}{c}
0 \\
0 \\
-\frac{1}{2} h^{\top} \Sigma_{\theta^{*}}^{-1} h
\end{array}\right)
,\left(\begin{array}{ccc}
\Sigma_{\tau} & C_{1}^{\top} \varphi_{\theta^{*}}^{-1} & -C_{1}^{\top} I_{\theta^{*}}^{-1} \varphi_{\theta^{*}} h \\
\varphi_{\theta^{*}}^{-1} C_{1} & \Sigma_{\theta^{*}} & -h \\
-h^{\top}\varphi_{\theta^{*}}I_{\theta^{*}}^{-1}C_{1} &-h& h^{\top} \Sigma_{\theta^{*}}^{-1} h
\end{array}\right)\right\},
$$
where  $C_{1}= Cov\left\{ \Psi(\theta_{n}),n^{1 / 2}\left(\hat{\tau}_{\mathrm{AMW}}^{\theta_{n}}-\tau^{\theta_{n}}\right)\right\}$. \\

\subsection{Martingale theory}
Define $D_{n}$ be a linear form for $Cov\left\{\Psi(\theta_{n}),n^{1 / 2}\left(\hat{\tau}_{\mathrm{AMW}}^{\theta_{n}}-\tau^{\theta_{n}}\right)\right\}=w_{1}^{\top}\Psi(\theta_{n})+w_{2}\left\{n^{1 / 2}\left(\hat{\tau}_{\mathrm{AMW}}^{\theta_{n}}-\tau^{\theta_{n}}\right)\right\}$ with weight $w_{1}=(w^{\top}_{11},w^{\top}_{12},w^{\top}_{13})^{\top},w_{2}$:

$$
\begin{aligned}
D_{n}&=w_{11}^{\top}n^{-1 / 2} \sum_{i=1}^{n}\left[\frac{\partial e\left(X_{i} , \alpha_{n}\right)}{\partial \alpha} \frac{A_{i}-e\left(X_{i} , \alpha_{n}\right)}{e\left(X_{i} , \alpha_{n}\right)\left\{1-e\left(X_{i} , \alpha_{n}\right)\right\}}\right]\\
&+w_{12}^{\top}n^{-1 / 2} \sum_{i=1}^{n}\left[\left(1-A_{i}\right) \frac{\partial u_{0}\left(X_{i} , \beta_{0, n}\right)}{\partial \beta_{0}}\left\{Y_{i}-u_{0}\left(X_{i} , \beta_{0, n}\right)\right\}\right]\\
&+w_{13}^{\top}n^{-1 / 2} \sum_{i=1}^{n}\left[A_{i} \frac{\partial u_{1}\left(X_{i} , \beta_{1, n}\right)}{\partial \beta_{1}}\left\{Y_{i}-u_{1}\left(X_{i} ,\beta_{1, n}\right)\right\}\right]\\
&+w_{2}n^{-1 / 2} \sum_{i=1}^{n}\left[\left\{u_{1}\left(X_{i},{\beta}_{1,n}\right)-u_{0}\left(X_{i}, {\beta}_{0,n}\right)\right\}-\tau^{\theta_{n}}\right]\\
&-w_{2}n^{-1 / 2}  \sum_{i=1}^{n}\left[(1-A_{i}) \frac{1}{\pi_{0}(e_{i,n})}\left\{Y_{i}-u_{0}\left(X_{i}, {\beta}_{0,n}\right)\right\}\right]\\
&+w_{2}n^{-1 / 2}  \sum_{i=1}^{n}\left[A_{i} \frac{1}{\pi_{1}(e_{i,n})}\left\{Y_{i}-u_{1}\left(X_{i}, {\beta}_{1,n}\right)\right\}\right]+o_{P}(1).
\end{aligned}
$$
\\
Hence, rewrite $D_{n}=\Sigma_{i'=1}\phi_{n,i'}$ into martingale presentation to implement martingale central limit theorem:\\
\\
For $0 \leq i' \leq n$,
$$
\phi^{1}_{n,i'}=w_{11}^{\top} n^{-1 / 2}\left[\frac{\partial e\left(X_{i'} , \alpha_{n}\right)}{\partial \alpha} \frac{e\left(X_{i'}\right)-e\left(X_{i'} , \alpha_{n}\right)}{e\left(X_{i'} , \alpha_{n}\right)\left\{1-e\left(X_{i'} , \alpha_{n}\right)\right\}}\right],
$$

$$
\phi^{2}_{n,i'}=w_{11}^{\top} n^{-1 / 2}\left[\frac{\partial e\left(X_{i'} , \alpha_{n}\right)}{\partial \alpha} \frac{A_{i'}-e\left(X_{i'}\right)}{e\left(X_{i'} , \alpha_{n}\right)\left\{1-e\left(X_{i'} , \alpha_{n}\right)\right\}}\right],
$$

$$
\phi^{3}_{n,i'}=w_{12}^{\top} n^{-1 / 2}\left\{1-e\left(X_{i'}\right)\right\} \frac{\partial u_{0}\left(X_{i'} ,\beta_{0, n}\right)}{\partial \beta_{0}}\left\{u_{0}\left(X_{i'}\right)-u_{0}\left(X_{i'} ,\beta_{0, n}\right)\right\},
$$

$$
\phi^{4}_{n,i'}=-w_{12}^{\top} n^{-1 / 2}\left\{A_{i'}-e\left(X_{i'}\right)\right\} \frac{\partial u_{0}\left(X_{i'} ,\beta_{0, n}\right)}{\partial \beta_{0}}\left\{u_{0}\left(X_{i'}\right)-u_{0}\left(X_{i'} ,\beta_{0, n}\right)\right\},
$$

$$
\phi^{5}_{n,i'}=w_{13}^{\top} n^{-1 / 2} e\left(X_{i'}\right) \frac{\partial u_{1}\left(X_{i'} ,\beta_{1, n}\right)}{\partial \beta_{1}}\left\{u_{1}\left(X_{i'}\right)-u_{1}\left(X_{i'} ,\beta_{1, n}\right)\right\},
$$

$$
\phi^{6}_{n,i'}=w_{13}^{\top} n^{-1 / 2}\left\{A_{i'}-e\left(X_{i'}\right)\right\} \frac{\partial u_{1}\left(X_{i'} ,\beta_{1, n}\right)}{\partial \beta_{1}}\left\{u_{1}\left(X_{i'}\right)-u_{1}\left(X_{i'} ,\beta_{1, n}\right)\right\},
$$

$$
\phi^{7}_{n,i'}=w_{2}n^{-1 / 2} \left[\left\{u_{1}\left(X_{i'}, {\beta}_{1,n}\right)-u_{0}\left(X_{i'}, {\beta}_{0,n}\right)\right\}-\tau^{\theta_{n}}\right],
$$

$$
\phi^{8}_{n,i'}=-w_{2}n^{-1 / 2}  (1-A_{i'}) \frac{1}{\pi_{0}(e_{i',n})}\left\{u_{0}(X_{i'})-u_{0}\left(X_{i'} ,\beta_{0, n}\right)\right\},
$$

$$
\phi^{9}_{n,i'}=w_{2}n^{-1 / 2}  A_{i'} \frac{1}{\pi_{1}(e_{i',n})}\left\{u_{1}(X_{i'})-u_{1}\left(X_{i'} ,\beta_{1, n}\right)\right\},
$$
\\
with $\sigma$ field like $\sigma_{i'}=\sigma\left(A_{1}, \ldots, A_{i'}, X_{1}, \ldots, X_{i'}\right)$.\\
\\
For $n < i' \leq 2n$,
$$
\phi^{10}_{n,i'}=w_{12}^{\top} n^{-1 / 2}\left\{1-A_{(i'-n)}\right\} \frac{\partial u_{0}\left(X_{i'} ,\beta_{0, n}\right)}{\partial \beta_{0}}\left[Y_{(i'-n)}-u_{0}\left\{X_{(i'-n)}\right\}\right],
$$

$$
\phi^{11}_{n,i'}=w_{13}^{\top} n^{-1 / 2}A_{(i'-n)} \frac{\partial u_{1}\left(X_{i'} ,\beta_{1, n}\right)}{\partial \beta_{1}}\left[Y_{(i'-n)}-u_{1}\left\{X_{(i'-n)}\right\}\right],
$$

$$
\phi^{12}_{n,i'}=-w_{2}n^{-1 / 2}  \left\{1-A_{(i'-n)}\right\} \frac{1}{\pi_{0}\left\{e_{(i'-n),n}\right\}}\left[Y_{(i'-n)}-u_{0}\left\{X_{(i'-n)}\right\}\right],
$$

$$
\phi^{13}_{n,i'}=w_{2}n^{-1 / 2}  A_{(i'-n)} \frac{1}{\pi_{1}\left\{e_{(i'-n),n}\right\}}\left[Y_{(i'-n)}-u_{1}\left\{X_{(i'-n)}\right\}\right],
$$
\\
with $\sigma$ field like $\sigma_{i'}=\sigma\left(A_{1}, \ldots, A_{n}, X_{1}, \ldots, X_{n},Y_{i'-1}, \ldots, Y_{i'-n} \right)$.
\\
It is easy to verify the martingale property based on given fields, then we can apply the central limitation theory as $D_{n} \sim N(0,V)$, where $V=\lim \Sigma^{2n}_{i'=1}\mathbb{E}(\phi^{2}_{n,i'}|\sigma_{i'})$. We derive the form of $C_{1}$, and $C_{1}=\left(c_{1}^{\top}, c_{2}^{\top}, c_{3}^{\top}\right)^{\top}$ with

$$
\begin{aligned}
c_{1}&=\mathbb{E}\left[\left\{u_{1}\left(X, \beta_{1}^{*}\right)-u_{0}\left(X, \beta_{0}^{*}\right)-\tau^{\theta^{*}}\right\}\frac{\partial e\left(X , \alpha^{*}\right)}{\partial \alpha} \frac{A-e\left(X, \alpha^{*}\right)}{e\left(X , \alpha^{*}\right)\left\{1-e\left(X, \alpha^{*}\right)\right\}}\right]\\
&+\mathbb{E}\left[\frac{1}{\pi_{0}(e^{{*}})}\left\{u_{0}(X)-u_{0}\left(X ,\beta_{0}^{*}\right)\right\}\frac{\partial e\left(X , \alpha^{*}\right)}{\partial \alpha} \frac{e\left(X, \alpha^{*}\right)}{e\left(X , \alpha^{*}\right)\left\{1-e\left(X , \alpha^{*}\right)\right\}}\right]\\
&+\mathbb{E}\left[\frac{1}{\pi_{1}(e^{{*}})}\left\{u_{1}(X)-u_{1}\left(X ,\beta_{1}^{*}\right)\right\}\frac{\partial e\left(X , \alpha^{*}\right)}{\partial \alpha} \frac{1-e\left(X, \alpha^{*}\right)}{e\left(X , \alpha^{*}\right)\left\{1-e\left(X , \alpha^{*}\right)\right\}}\right],
\end{aligned}
$$

$$
\begin{aligned}
c_{2}&=\mathbb{E}\left[\left\{u_{1}\left(X, \beta_{1}^{*}\right)-u_{0}\left(X, \beta_{0}^{*}\right)-\tau^{\theta^{*}}\right\}(1-A)\frac{\partial u_{0}\left(X ,\beta_{0}^{*}\right)}{\partial \beta_{0}}\left\{u_{0}\left(X\right)-u_{0}\left(X ,\beta_{0}^{*}\right)\right\}\right]\\
&-\mathbb{E}\left[ \frac{\partial u_{0}\left(X , \beta_{0}^{*}\right)}{\partial \beta_{0}}\left\{u_{0}(X)-u_{0}\left(X , \beta_{0}^{*}\right)\right\}^{2} \frac{1}{\pi_{0}(e^{{*}})}\right]\\
&-\mathbb{E}\left\{ \frac{\partial u_{0}\left(X , \beta_{0}^{*}\right)}{\partial \beta_{0}} \sigma^{2}_{0}\frac{1}{\pi_{0}(e^{{*}})}\right\},
\end{aligned}
$$

$$
\begin{aligned}
c_{3}&=\mathbb{E}\left[\left\{u_{1}\left(X, \beta_{1}^{*}\right)-u_{0}\left(X, \beta_{0}^{*}\right)-\tau^{\theta^{*}}\right\}A\frac{\partial u_{1}\left(X ,\beta_{1}^{*}\right)}{\partial \beta_{1}}\left\{u_{1}\left(X\right)-u_{1}\left(X ,\beta_{1}^{*}\right)\right\}\right]\\
&+\mathbb{E}\left[ \frac{\partial u_{1}\left(X , \beta_{1}^{*}\right)}{\partial \beta_{1}}\left\{u_{1}(X)-u_{1}\left(X , \beta_{1}^{*}\right)\right\}^{2} \frac{1}{\pi_{1}(e^{{*}})}\right]\\
&+\mathbb{E}\left\{ \frac{\partial u_{1}\left(X , \beta_{1}^{*}\right)}{\partial \beta_{1}} \sigma^{2}_{1}\frac{1}{\pi_{1}(e^{*})}\right\}.
\end{aligned}
$$

\subsection{Le Cam' third theory}
Abstract Le Cam' third theory from van der Vaart(2000):\\
If
$$
\left(X_{n}, \log \frac{d Q_{n}}{d P_{n}}\right) {\overset{P_{n}}{\rightarrow}} N_{k+1}\left\{\left(\begin{array}{c}
u \\
-\frac{1}{2} \sigma^{2}
\end{array}\right),\left(\begin{array}{cc}
\Sigma & \tau \\
\tau^{T} & \sigma^{2}
\end{array}\right)\right\},
$$
then 
$$
X_{n} {\overset{Q_{n}}{\rightarrow}} N_{k}(u+\tau, \Sigma).
$$
\\
Le cam' third theory is applied to get asymptotic distribution for AWM estimator. Firstly, specific form of $\tau^{\theta_{n}}$ is provided with AIPW estimator, since AIPW is related with both propensity scores and prognostic scores:
$$\tau^{\theta_{n}}=\mathbb{E}\left[u_{1}\left(X, \beta_{1,n}\right)-u_{0}\left(X, \beta_{0,n}\right)+\left\{ \frac{A{R}_{n}}{{e}(X,\alpha_{n})}-\frac{(1-A){R}_{n}}{1-{e}(X,\alpha_{n})}\right\}\right].$$
\\
Then, do Taylor expansion:
$$\tau^{\theta_{n}}=\tau^{\theta^{*}}+\frac{\partial \tau^{\theta}} {\partial  \theta }|_{\theta=\theta^{*}}(\theta^{*}-\theta_{n})+o(n^{-1/2}).$$
What's more, denote$(\frac{\partial \tau^{\theta}} {\partial  \theta }|_{\theta=\theta^{*}})=C_{2}$ under distribution of $\theta^{*}$ apply  Le Cam' third lemma to get:
$$
\left\{\begin{array}{c}
n^{1 / 2}\left(\hat{\tau}_{\mathrm{AMW}}^{\theta_{n}}-\tau\right) \\
n^{1 / 2}\left(\hat{\theta}-\theta_{n}\right)
\end{array}\right\} \rightarrow \mathcal{N}\left\{\left(\begin{array}{cc}
-h^{\top}\varphi_{\theta^{*}}I_{\theta^{*}}^{-1}C_{1}- C_{2}^{\top}h\\
-h
\end{array}\right),\left(\begin{array}{cc}
\Sigma_{\tau} & C_{1}^{\top} \varphi_{\theta^{*}}^{-1} \\
\varphi_{\theta^{*}}^{-1} C_{1} & \Sigma_{\theta^{*}}
\end{array}\right)\right\}.
$$
By replacing $\theta_{n}$  with $\theta^{*}$,
$$
\left\{\begin{array}{c}
n^{1 / 2}\left(\hat{\tau}_{\mathrm{AMW}}^{\theta^{*}+ h/\sqrt{n}}-\tau\right) \\
n^{1 / 2}\left(\hat{\theta}-\theta^{*}\right)
\end{array}\right\} \rightarrow \mathcal{N}\left\{\left(\begin{array}{cc}
-h^{\top}\varphi_{\theta^{*}}I_{\theta^{*}}^{-1}C_{1}- C_{2}^{\top}h \\
0
\end{array}\right),\left(\begin{array}{cc}
\Sigma_{\tau} & C_{1}^{\top} \varphi_{\theta^{*}}^{-1} \\
\varphi_{\theta^{*}}^{-1} C_{1} & \Sigma_{\theta^{*}}
\end{array}\right)\right\}.
$$
Apply the conditional normal  distribution:
$$n^{1 / 2}\left(\hat{\tau}_{\mathrm{AMW}}^{\hat{\theta}}-\tau\right)|n^{1 / 2}\left(\hat{\theta}-\theta^{*}\right)=h \rightarrow N\left(C_{2}^{\top}h, \Sigma_{\tau}-C_{1}^{\top} I_{\theta^{*}}^{-1} C_{1}\right).$$
Marginalize over the $h$ to get unconditional distribution:
$$n^{1 / 2}\left(\hat{\tau}_{\mathrm{AMW}}^{\theta^{*}+ h/\sqrt{n}}-\tau\right) \rightarrow N\left(0, \Sigma_{\tau}-C_{1}^{\top} I_{\theta^{*}}^{-1} C_{1}+C_{2}^{\top}\Sigma_{\theta^{*}}C_{2}\right).$$

\section{Proof for Theorem 4}
The idea is identical to proof for Theorem 3, and we just give the main steps to derive Theorem 4. 

\subsection{Asymptotic joint distribution}
Under distribution of $\theta_{n}$, the joint distribution can be derived as:

\[
\left\{
\begin{array}{c}
n^{1/2}\left(\hat{\tau}_{\mathrm{AMW}}^{t,\theta_{n}} - \tau^{t,\theta_{n}}\right) \\
n^{1/2}\left(\hat{\theta} - \theta_{n}\right) \\
\log\left( {p}^{\theta^{*}} /  p^{\theta_{n}} \right)
\end{array}
\right\}
\overset{d}{\rightarrow}
\mathcal{N} \left(
\left(
\begin{array}{c}
0 \\
0 \\
-\frac{1}{2} h^{\top} \Sigma_{\theta^{*}}^{-1} h
\end{array}
\right),
\left(
\begin{array}{ccc}
\Sigma^{t}_{\tau} & C^{t\top}_{1} \varphi_{\theta^{*}}^{-1} & -C^{t\top}_{1} I_{\theta^{*}}^{-1} \varphi_{\theta^{*}} h \\
\varphi_{\theta^{*}}^{-1} C^{t}_{1} & \Sigma_{\theta^{*}} & -h \\
- h^{\top} \varphi_{\theta^{*}} I_{\theta^{*}}^{-1} C^{t}_{1} & -h & h^{\top} \Sigma_{\theta^{*}}^{-1} h
\end{array}
\right)
\right)
\]
where  $C^{t}_{1}= Cov\left\{ \Psi(\theta_{n}),n^{1 / 2}\left(\hat{\tau}_{\mathrm{AMW}}^{t,\theta_{n}}-\tau^{t,\theta_{n}}\right)\right\}$. \\

\subsection{Martingale theory}
Define $D_{n}$ be a linear form for $Cov\left\{\Psi(\theta_{n}),n^{1 / 2}\left(\hat{\tau}_{\mathrm{AMW}}^{t,\theta_{n}}-\tau^{t,\theta_{n}}\right)\right\}=w_{1}^{ t,\top}\Psi(\theta_{n})+w^{t}_{2}\left\{n^{1 / 2}\left(\hat{\tau}_{\mathrm{AMW}}^{t,\theta_{n}}-\tau^{t,\theta_{n}}\right)\right\}$ with weight $w_{1}^{t}=(w^{ t,\top}_{11},w^{ t,\top}_{12},w^{ t,\top}_{13})^{\top},w^{t}_{2}$:

$$
\begin{aligned}
D^{t}_{n}&=w^{ t,\top}_{11}n_{1}^{-1 / 2} \sum_{i=1}^{n}\left[\frac{\partial e\left(X_{i} , \alpha_{n}\right)}{\partial \alpha} \frac{A_{i}-e\left(X_{i} , \alpha_{n}\right)}{e\left(X_{i} , \alpha_{n}\right)\left\{1-e\left(X_{i} , \alpha_{n}\right)\right\}}\right]\\
&+w^{ t,\top}_{12}n_{1}^{-1 / 2} \sum_{i=1}^{n}\left[\left(1-A_{i}\right) \frac{\partial u_{0}\left(X_{i} , \beta_{0, n}\right)}{\partial \beta_{0}}\left\{Y_{i}-u_{0}\left(X_{i} , \beta_{0, n}\right)\right\}\right]\\
&+w^{ t,\top}_{13}n_{1}^{-1 / 2} \sum_{i=1}^{n}\left[A_{i} \frac{\partial u_{1}\left(X_{i} , \beta_{1, n}\right)}{\partial \beta_{1}}\left\{Y_{i}-u_{1}\left(X_{i} , \beta_{1, n}\right)\right\}\right]\\
&+w^{t}_{2}n_{1}^{-1 / 2} \sum_{i=1}^{n}\left[\left\{Y_{i}-u_{0}\left(X_{i}, {\beta}_{0,n}\right)\right\}A_{i}-\tau^{t,\theta_{n}}\right]\\
&-w^{t}_{2}n_{1}^{-1 / 2}  \sum_{i=1}^{n}\left[(1-A_{i}) \frac{1}{\pi_{0}(e_{i,n})}\left\{Y_{i}-u_{0}\left(X_{i}, {\beta}_{0,n}\right)\right\}\right]\\
&+w^{t}_{2}n_{1}^{-1 / 2}  \sum_{i=1}^{n}\left[A_{i}\left\{Y_{i}-u_{1}\left(X_{i}, {\beta}_{1,n}\right)\right\}\right]+o_{P}(1).
\end{aligned}
$$
\\

Apply martingale theory, for $0 \leq i' \leq n$,
$$
\phi^{t,1}_{n,i'}=w_{11}^{ t,\top} n_{1}^{-1 / 2}\left[\frac{\partial e\left(X_{i'} , \alpha_{n}\right)}{\partial \alpha} \frac{e\left(X_{i'}\right)-e\left(X_{i'} , \alpha_{n}\right)}{e\left(X_{i'} , \alpha_{n}\right)\left\{1-e\left(X_{i'} , \alpha_{n}\right)\right\}}\right],
$$

$$
\phi^{t,2}_{n,i'}=w_{11}^{ t,\top} n_{1}^{-1 / 2}\left[\frac{\partial e\left(X_{i'} , \alpha_{n}\right)}{\partial \alpha} \frac{A_{i'}-e\left(X_{i'}\right)}{e\left(X_{i'} , \alpha_{n}\right)\left\{1-e\left(X_{i'} , \alpha_{n}\right)\right\}}\right],
$$

$$
\phi^{t,3}_{n,i'}=w_{12}^{ t,\top} n_{1}^{-1 / 2}\left\{1-e\left(X_{i'}\right)\right\} \frac{\partial u_{0}\left(X_{i'} ,\beta_{0, n}\right)}{\partial \beta_{0}}\left\{u_{0}\left(X_{i'}\right)-u_{0}\left(X_{i'} ,\beta_{0, n}\right)\right\},
$$

$$
\phi^{t,4}_{n,i'}=-w_{12}^{ t,\top} n_{1}^{-1 / 2}\left\{A_{i'}-e\left(X_{i'}\right)\right\} \frac{\partial u_{0}\left(X_{i'} ,\beta_{0, n}\right)}{\partial \beta_{0}}\left\{u_{0}\left(X_{i'}\right)-u_{0}\left(X_{i'} ,\beta_{0, n}\right)\right\},
$$

$$
\phi^{t,5}_{n,i'}=w_{13}^{ t,\top} n_{1}^{-1 / 2} e\left(X_{i'}\right) \frac{\partial u_{1}\left(X_{i'} ,\beta_{1, n}\right)}{\partial \beta_{1}}\left\{u_{1}\left(X_{i'}\right)-u_{1}\left(X_{i'} ,\beta_{1, n}\right)\right\},
$$

$$
\phi^{t,6}_{n,i'}=w_{13}^{ t,\top} n_{1}^{-1 / 2}\left\{A_{i'}-e\left(X_{i'}\right)\right\} \frac{\partial u_{1}\left(X_{i'} ,\beta_{1, n}\right)}{\partial \beta_{1}}\left\{u_{1}\left(X_{i'}\right)-u_{1}\left(X_{i'} ,\beta_{1, n}\right)\right\},
$$

$$
\phi^{t,7}_{n,i'}=w^{t}_{2}n_{1}^{-1 / 2} \left[\left\{u\left(X_{i'}, {\beta}_{1,n}\right)-u\left(X_{i'}, {\beta}_{0,n}\right)\right\}A_{i'}-\tau^{t,\theta_{n}}\right],
$$

$$
\phi^{t,8}_{n,i'}=w^{t}_{2}n_{1}^{-1 / 2}  A_{i'} \left\{u_{1}(X_{i'})-u_{1}\left(X_{i'} ,\beta_{1, n}\right)\right\},
$$

$$
\phi^{t,9}_{n,i'}=w^{t}_{2}n_{1}^{-1 / 2} \left\{u_{1}\left(X_{i'}, {\beta}_{1,n}\right)-u_{0}\left(X_{i'}, {\beta}_{0,n}\right)\right\}A_{i'}-\tau^{\theta_{n}},
$$
\\
with $\sigma$ field like $\sigma_{i'}=\sigma\left(A_{1}, \ldots, A_{i'}, X_{1}, \ldots, X_{i'}\right)$.\\
\\
For $n < i' \leq 2n$,
$$
\phi^{t,10}_{n,i'}=w^{t}_{2}n_{1}^{-1 / 2} \left[\left\{Y_{(i'-n)}-u_{1}\left(X_{(i'-n)}, {\beta}_{1,n}\right)\right\}A_{(i'-n)}\right],
$$

$$
\phi^{t,11}_{n,i'}=w_{12}^{ t,\top} n_{1}^{-1 / 2}\left\{1-A_{(i'-n)}\right\} \frac{\partial u_{0}\left(X_{(i'-n)} ,\beta_{0, n}\right)}{\partial \beta_{0}}\left[Y_{(i'-n)}-u_{0}\left\{X_{(i'-n)}\right\}\right],
$$

$$
\phi^{t,12}_{n,i'}=w_{13}^{ t,\top} n_{1}^{-1 / 2}A_{(i'-n)} \frac{\partial u_{1}\left(X_{(i'-n)} ,\beta_{1, n}\right)}{\partial \beta_{1}}\left[Y_{(i'-n)}-u_{1}\left\{X_{(i'-n)}\right\}\right],
$$

$$
\phi^{t,13}_{n,i'}=-w^{t}_{2}n_{1}^{-1 / 2}  \left\{1-A_{(i'-n)}\right\} \frac{1}{\pi_{0}\left\{e_{(i'-n),n}\right\}}\left[Y_{(i'-n)}-u_{0}\left\{X_{(i'-n)}\right\}\right],
$$

$$
\phi^{t,14}_{n,i'}=w^{t}_{2}n_{1}^{-1 / 2}  A_{(i'-n)} \left[Y_{(i'-n)}-u_{1}\left\{X_{(i'-n)}\right\}\right],
$$
with $\sigma$ field like $\sigma_{i'}=\sigma\left(A_{1}, \ldots, A_{n}, X_{1}, \ldots, X_{n},Y_{i'-1}, \ldots, Y_{i'-n} \right)$.
\\
And we obtain $C_{1}^{t}$ with:
$$
\begin{aligned}
c^{t}_{1}&=\mathbb{E}\left[\left\{u(X, \beta_{1}^{*})A-u\left(X, \beta_{0}^{*}\right)A-\tau^{t,\theta^{*}}\right\}\frac{\partial e\left(X , \alpha^{*}\right)}{\partial \alpha} \frac{A-e\left(X, \alpha^{*}\right)}{e\left(X , \alpha^{*}\right)\left\{1-e\left(X, \alpha^{*}\right)\right\}}\right]\\
&+\mathbb{E}\left[\frac{1}{\pi_{0}(e^{{*}})}\left\{u_{0}(X)-u_{0}\left(X ,\beta_{0}^{*}\right)\right\}\frac{\partial e\left(X , \alpha^{*}\right)}{\partial \alpha} \frac{e\left(X, \alpha^{*}\right)}{e\left(X , \alpha^{*}\right)\left\{1-e\left(X , \alpha^{*}\right)\right\}}\right]\\
&+\mathbb{E}\left[\left\{u_{1}(X)-u_{1}\left(X ,\beta_{1}^{*}\right)\right\}\frac{\partial e\left(X , \alpha^{*}\right)}{\partial \alpha} \frac{1-e\left(X, \alpha^{*}\right)}{e\left(X , \alpha^{*}\right)\left\{1-e\left(X , \alpha^{*}\right)\right\}}\right],
\end{aligned}
$$

$$
\begin{aligned}
c^{t}_{2}&=\mathbb{E}\left[\left\{u\left(X, \beta_{1}^{*}\right)A-u\left(X, \beta_{0}^{*}\right)A-\tau^{t,\theta^{*}}\right\}(1-A)\frac{\partial u_{0}\left(X ,\beta_{0}^{*}\right)}{\partial \beta_{0}}\left\{u_{0}\left(X\right)-u_{0}\left(X ,\beta_{0}^{*}\right)\right\}\right]\\
&-\mathbb{E}\left[ \frac{\partial u_{0}\left(X , \beta_{0}^{*}\right)}{\partial \beta_{0}}\left\{u_{0}(X)-u_{0}\left(X , \beta_{0}^{*}\right)\right\}^{2} \frac{1}{\pi_{0}(e^{{*}})}\right]\\
&-\mathbb{E}\left\{ \frac{\partial u_{0}\left(X , \beta_{0}^{*}\right)}{\partial \beta_{0}} \sigma^{2}_{0}\frac{1}{\pi_{0}(e^{{*}})}\right\},
\end{aligned}
$$

$$
\begin{aligned}
c^{t}_{3}=&\mathbb{E}\left[\left\{u\left(X, \beta_{1}^{*}\right)A-u\left(X, \beta_{0}^{*}\right)A-\tau^{t,\theta^{*}}\right\}A\frac{\partial u_{1}\left(X ,\beta_{1}^{*}\right)}{\partial \beta_{1}}\left\{u_{1}\left(X\right)-u_{1}\left(X ,\beta_{1}^{*}\right)\right\}\right]\\
+&\mathbb{E}\left[ \frac{\partial u_{1}\left(X , \beta_{1}^{*}\right)}{\partial \beta_{1}}\left\{u_{1}(X)-u_{1}\left(X , \beta_{1}^{*}\right)\right\}^{2} \right]\\
+&\mathbb{E}\left[\left\{Y-u_{1}(X)\right\}\frac{\partial u_{1}\left(X , \beta_{1}^{*}\right)}{\partial \beta_{1}} \left\{Y-u_{1}\left(X , \beta_{1}^{*}\right)\right\}\right]\\
+&\mathbb{E}\left\{ \frac{\partial u_{1}\left(X , \beta_{1}^{*}\right)}{\partial \beta_{1}} \sigma^{2}_{1}\right\}.
\end{aligned}
$$

\subsection{Le Cam' third theory}

$$\tau^{t,\theta_{n}}=\mathbb{E}\left[YA-u_{0}\left(X, \beta_{0,n}\right)A+\left\{ A{R_{n}}-\frac{(1-A)e(X,\alpha_{n})R_{n}}{1-{e}(X,\alpha_{n})}\right\} \right].$$
\\
By Taylor expansion:
$$\tau^{t,\theta_{n}}=\tau^{t,\theta^{*}}+\frac{\partial \tau^{t,\theta}} {\partial  \theta }|_{\theta=\theta^{*}}(\theta^{*}-\theta_{n})+o(n^{-1/2}).$$
\\
Marginalize over the $h$ to get unconditional distribution:
\[
n^{1/2}\left(\hat{\tau}_{\mathrm{AMW}}^{t,\theta^{*}+ h/\sqrt{n}} - \tau^{t}\right) 
\rightarrow 
\mathcal{N}\left(0,\, \Sigma^{t}_{\tau} 
- C^{t\top}_{1} I_{\theta^{*}}^{-1} C^{t}_{1} 
+ C^{t\top}_{2} \Sigma_{\theta^{*}} C^{t}_{2} \right).
\]

\section{Tables}

\begin{table}[H] 
\caption{Standard difference for NSW-DS data regarding ATE}
\centering
\resizebox{\columnwidth}{!}{%
\begin{tabular}{|l|lllllll|} \hline covirates    & age   & educ  & black & hisp   & married & nodegr & re75   \\ before & 0.107 & 0.144 & 0.044 & -0.170 & 0.094   & -0.306 & 0.084  \\  after  & 0.014 & 0.046 & 0.069 & -0.090 & -0.001   & -0.016 & -0.062 \\ \hline \end{tabular}%
}
\label{std_Labor} 
\end{table}

\begin{table}[H] 
\caption{Standard difference for NSW-DS data regarding ATT}
\centering
\resizebox{\columnwidth}{!}{%
\begin{tabular}{|l|lllllll|} \hline covirates    & age   & educ  & black & hisp   & married & nodegr & re75   \\  before  & 0.107 & 0.144 & 0.044 & -0.170 & 0.094   & -0.306 & 0.084  \\ after   & 0.013 & 0.046 & 0.069 & -0.092 & -0.003   & -0.016 & -0.062 \\ \hline \end{tabular}%
}
\label{std_LaboraATT} 
\end{table}

\begin{table}[H] 
\caption{Standard difference for ACGT175 data regarding ATE}
\centering
\resizebox{\columnwidth}{!}{%
\begin{tabular}{|l|llllllll|} \hline covirates    & age     & wtkg   & hemo   & homo   & drugs & karnof  & oprior & z30   \\  before  & 0.036   & -0.090 & 0.017  & 0.008  & 0.072 & 0.084   & 0.010  & 0.066 \\  after   & 0.016   & 0.004  & 0.007  & -0.003 & 0.006 & -0.002  & 0.003  & 0.000 \\ \hline covirates    & preanti & race   & gender & str2   & strat & symptom & cd40   & cd80  \\  before  & 0.027   & -0.042 & -0.001 & 0.085  & 0.072 & 0.048   & -0.022 & 0.012 \\  after   & 0.008   & -0.014 & -0.004 & 0.005  & 0.006 & 0.005   & -0.012 & 0.008 \\ \hline 
\end{tabular} %
}
\label{std_AIDS} 
\end{table}

\begin{table}[]
\centering
\caption{Standard difference for RHC data regarding ATT}
\resizebox{\columnwidth}{!}{%
\begin{tabular}{|l|lllll|}
\hline
covariate & age         & sexMale   & raceother    & racewhite & edu              \\
before    & -0.061      & 0.093     & 0.021        & 0.015     & 0.091            \\
after     & 0.003       & 0.033     & -0.021       & 0.018     & -0.008           \\ \hline
covariate & income      & income    & income       & ninsclas  & nonsclasMedicare \\
          & \$11-\$25k    & \$25-\$50k  & Under \$11K  & Medicare  & \&Medicaid       \\
before    & 0.015       & 0.108     & -0.124       & -0.075    & -0.058           \\
after     & -0.012      & -0.003    & 0.021        & -0.019    & 0.027            \\ \hline
covariate & ninsclas No & ninsclas  & ninsclas     & cat1CHF   & cat1Cirrhosis    \\
          & insurance   & Private   & Private\&Med &           &                  \\
before    & 0.043       & 0.137     & 0.035        & 0.097     & -0.139           \\
after     & 0           & 0.002     & -0.002       & 0.006     & -0.013           \\ \hline
covariate & cat1Colon   & cat1Coma  & cat1COPD     & cat1Lung  & cat1MOSF         \\
          & Cancer      &           &              & Cancer    & w/Malignancy     \\
before    & -0.035      & -0.198    & -0.317       & -0.089    & 0.018            \\
after     & -0.014      & -0.016    & -0.015       & -0.041    & 0.001            \\ \hline
covariate & cat1MOSF    & das2d3pc  & dnr1Yes      & caNo      & caYes            \\
          & w/Sepsis    &           &              &           &                  \\
before    & 0.42        & 0.062     & -0.219       & 0.103     & -0.071           \\
after     & 0.009       & 0.001     & -0.04        & 0.012     & -0.003           \\ \hline
covariate & surv2md1    & aps1      & scoma1       & wtkilo1   & temp1            \\
before    & -0.198      & 0.491     & -0.108       & 0.252     & -0.021           \\
after     & -0.011      & 0.03      & -0.024       & 0.043     & -0.017           \\ \hline
covariate & meanbp1     & resp1     & hrt1         & pafi1     & paco21           \\
before    & -0.438      & -0.165    & 0.147        & -0.419    & -0.24            \\
after     & 0.009       & 0.019     & 0.024        & -0.003    & -0.003           \\ \hline
covariate & ph1         & wblc1     & hema1        & sod1      & pot1             \\
before    & -0.12       & 0.084     & -0.262       & -0.092    & -0.027           \\
after     & 0.01        & 0.084     & -0.008       & -0.019    & -0.016           \\ \hline
covariate & crea1       & bili1     & alb1         & respYes   & cardYes          \\
before    & 0.268       & 0.148     & -0.237       & -0.265    & 0.295            \\
after     & 0.005       & -0.014    & 0.018        & -0.006    & 0.021            \\ \hline
covariate & neuroYes    & gastrYes  & renalYes     & metaYes   & hemaYes          \\
before    & -0.331      & 0.122     & 0.119        & -0.028    & -0.061           \\
after     & -0.016      & -0.005    & 0.023        & 0.007     & 0.004            \\ \hline
covariate & sepsYes     & traumaYes & orthoYes     & cardiohx  & chfhx            \\
before    & 0.238       & 0.111     & 0.028        & 0.117     & 0.07             \\
after     & 0.008       & 0.005     & 0.002        & 0.004     & 0.005            \\ \hline
covariate & dementhx    & psychhx   & chrpulhx     & renalhx   & liverhx          \\
before    & -0.158      & -0.139    & -0.188       & 0.032     & -0.048           \\
after     & -0.036      & 0.002     & 0.022        & 0.018     & -0.004           \\ \hline
covariate & gibledhx    & malighx   & immunhx      & transhx   & amihx            \\
before    & -0.069      & -0.101    & 0.081        & 0.173     & 0.076            \\
after     & -0.021      & -0.011    & 0.001        & 0.016     & -0.004           \\ \hline
\end{tabular}
}
\end{table}

\section{Simulation}
\subsection{AIDS Clinical Trials Group Study }
\citet{hammer1996trial} designed a double-blind study to assess the
efficacy of AIDS treatments, comparing the effects of a single treatment
of either zidovudine or didanosine to the combination treatment of
zidovudine plus didanosine or zalcitabine. The dataset used for this
study, AIDS Clinical Trials Group Study 175 (ACTG175), is available in the R package \textquotedbl speff2trial.\textquotedbl{} As HIV
can attack the immune system and reduce CD4 cell counts, the primary
outcome of interest in our design is the difference between CD4 cell
counts after 96 weeks and the baseline. Units with missing outcomes
are excluded to maintain the validity of the modeling process, resulting
in a reduced model with 1342 individuals. While the original study
was designed as a randomized experiment, missing values posed challenges
when comparing the two therapy groups directly. Thus, it is necessary to employ the AMW estimator to eliminate confounding bias and obtain
reliable effects. The outcome model utilizes a linear model with all
14 covariates. 
Figure 3 displays the standardized differences for covariates
before and after matching, which demonstrates that the AMW estimator
effectively enhances the balance between the two groups. The average
treatment effect (ATE) is calculated to be 40.852 with a standard
error of 8.220, yielding a $95\%$ empirical lower bound for ATE of
24.740. This result, which is greater than zero, suggests a significant
improvement in CD4 cell counts resulting from combination therapy.
\subsection{Right Heart Catheterization}

\citet{connors1996effectiveness} and \citet{hirano2001estimation}
analyzed the RHC dataset to investigate the safety and effectiveness
of the Right Heart Catheterization procedure for severely ill patients
in ICUs. The RHC procedure was previously believed to increase patient
survival rates, but subsequent reanalysis by \citet{hirano2001estimation}
suggested otherwise. We aim to use the AMW estimator to analyze the
RHC dataset and evaluate its effectiveness on this challenging dataset.
The RHC dataset, which contains 5735 units with 63 variables, is available
at https://hbiostat.org/data/. For our analysis, we select 50 variables
and convert categorical variables into dummy variables for both the
propensity score and outcome models. We model the outcome variable,
which is a binary indicator of whether the patient survives for 30
days (the survival patient denoted as ``1'', and the dead patient
denoted as ``0''), using logistic regression. Our goal is to estimate
the causal risk difference by comparing the survival probabilities
between the two experimental groups.

Figure 4 demonstrates that the use of the AMW estimator
results in reduced differences between the observed covariates in
the two groups before and after matching. Our estimated ATT of -0.0641
is consistent with the effects obtained by overlap and optimal matching
estimators, as reported by \citet{li2016balancing}. However, the
standard error for our estimator, 0.0241, is higher than those obtained
by the other two estimators. Our findings suggest that the RHC procedure
is associated with increased risk for patients, which contrasts with
the effects estimated by traditional approaches that do not account
for confounding bias.

\bibliographystyle{tfcad}
\bibliography{Ch2}